\documentclass[conference]{IEEEtran}
\IEEEoverridecommandlockouts
\usepackage{caption}
\usepackage{algorithm}
\usepackage[algo2e,noend]{algorithm2e}

\usepackage[hyphens]{url} %To break URLs
\usepackage{cite}
\usepackage{amsmath,amssymb,amsfonts}
\usepackage{algorithmic}
\usepackage{graphicx}
\usepackage{textcomp}
\usepackage{xcolor}
\usepackage{makecell} % Add this line to use the makecell package

\usepackage[bookmarks=true,breaklinks=true,colorlinks,citecolor=blue,linkcolor=blue,urlcolor=blue]{hyperref}

\def\BibTeX{{\rm B\kern-.05em{\sc i\kern-.025em b}\kern-.08em
    T\kern-.1667em\lower.7ex\hbox{E}\kern-.125emX}}
\begin{document}

\title{Pushing the Performance Envelope of DNN-based
Recommendation Systems Inference on GPUs\\
% {\footnotesize \textsuperscript{*}Note: Sub-titles are not captured for https://ieeexplore.ieee.org  and
% should not be used}
% \thanks{Identify applicable funding agency here. If none, delete this.}
}

\author{
\IEEEauthorblockN{\hspace{-4mm}Rishabh Jain}
\IEEEauthorblockA{\textit{Computer Science and Engineering} \\
\textit{The Pennsylvania State University}\\
University Park, PA, USA \\
rishabh@psu.edu}
\and
\IEEEauthorblockN{\hspace{-0.5in}Vivek M. Bhasi}
\IEEEauthorblockA{\textit{\hspace{-0.3in}Computer Science and Engineering} \\
\textit{\hspace{-0.3in}The Pennsylvania State University}\\
\hspace{-0.3in}University Park, PA, USA \\
\hspace{-0.3in}vmbhasi@psu.edu}
\and
\IEEEauthorblockN{Adwait Jog}
\IEEEauthorblockA{\textit{Computer Science} \\
\textit{University of Virginia}\\
Charlottesville, VA, USA \\
ajog@virginia.edu}
\and
\IEEEauthorblockN{\hspace{0.5in}Anand Sivasubramaniam}
\IEEEauthorblockA{\textit{\hspace{0.5in}Computer Science and Engineering} \\
\textit{\hspace{0.5in}The Pennsylvania State University}\\
\hspace{0.5in}University Park, PA, USA \\
\hspace{0.5in}axs53@psu.edu}
\and
\IEEEauthorblockN{\hspace{0.3in} Mahmut T. Kandemir}
\IEEEauthorblockA{\textit{\hspace{0.3in}Computer Science and Engineering} \\
\textit{\hspace{0.3in}The Pennsylvania State University}\\
\hspace{0.3in}University Park, PA, USA \\
\hspace{0.3in}mtk2@psu.edu}
\and
\IEEEauthorblockN{\hspace{0.15in}Chita R. Das}
\IEEEauthorblockA{\textit{\hspace{0.15in}Computer Science and Engineering} \\
\textit{\hspace{0.15in}The Pennsylvania State University}\\
\hspace{0.15in}University Park, PA, USA \\
\hspace{0.15in}cxd12@psu.edu}
}

% \author{
% \IEEEauthorblockN{Rishabh Jain}
% \IEEEauthorblockA{\textit{Computer Science and Engineering} \\
% \textit{The Pennsylvania State University}\\
% University Park, PA, USA \\
% rishabh@psu.edu}
% \and
% \IEEEauthorblockN{Vivek M. Bhasi}
% \IEEEauthorblockA{\textit{Computer Science and Engineering} \\
% \textit{The Pennsylvania State University}\\
% University Park, PA, USA \\
% vmbhasi@psu.edu}
% \and
% \IEEEauthorblockN{Adwait Jog}
% \IEEEauthorblockA{\textit{Computer Science} \\
% \textit{University of Virginia}\\
% Charlottesville, VA, USA \\
% ajog@virginia.edu}
% \and
% \IEEEauthorblockN{Anand Sivasubramaniam}
% \IEEEauthorblockA{\textit{Computer Science and Engineering} \\
% \textit{The Pennsylvania State University}\\
% University Park, PA, USA \\
% axs53@psu.edu}
% \and
% \IEEEauthorblockN{Mahmut T. Kandemir}
% \IEEEauthorblockA{\textit{Computer Science and Engineering} \\
% \textit{The Pennsylvania State University}\\
% University Park, PA, USA \\
% mtk2@psu.edu}
% \and
% \IEEEauthorblockN{Chita R. Das}
% \IEEEauthorblockA{\textit{Computer Science and Engineering} \\
% \textit{The Pennsylvania State University}\\
% University Park, PA, USA \\
% cxd12@psu.edu}
% }

\maketitle

\begin{abstract}
Personalized recommendation is a ubiquitous application on the internet, with many industries and hyperscalers extensively leveraging Deep Learning Recommendation Models (DLRMs) for their personalization needs (like ad serving or movie suggestions). With growing model and dataset sizes pushing computation and memory requirements, GPUs are being increasingly preferred for executing DLRM inference. However, serving newer DLRMs, while meeting acceptable latencies, continues to remain challenging, making traditional deployments increasingly more GPU-hungry, resulting in higher inference serving costs. In this paper, we show that the embedding stage continues to be the primary bottleneck in the GPU inference pipeline, leading up to a 3.2$\times$ embedding-only performance slowdown. 

To thoroughly grasp the problem, we conduct a detailed \emph{microarchitecture characterization} and highlight the presence of low occupancy in the standard embedding kernels. By leveraging direct compiler optimizations, we achieve optimal occupancy, pushing the performance by up to 53\%. 
% \adwait{are you doing any compiler optimization?}
Yet, long memory latency stalls continue to exist. To tackle this challenge, we propose specialized plug-and-play-based \emph{software prefetching and L2 pinning techniques}, which help in hiding and decreasing the latencies. Further, we propose combining them, as they complement each other. Experimental evaluations using A100 GPUs with large models and datasets show that our proposed techniques improve performance by up to 103\% for the embedding stage, and up to 77\% for the overall DLRM inference pipeline.  
\end{abstract}


\vspace{-0.1in}
\section{Introduction}
\label{sec:introduction}

%In recent years, AI has rapidly advanced, thanks to innovations across multiple fronts \cite{nebula, LLM_survey, stash, skipper, arxiv_stash}, transforming numerous industries by powering a wide range of internet applications. Recommendation Systems, in particular, are 

Recommendation Systems are the driving force for many internet applications such as social networks~\cite{fb_recsys,instagram_recsys, tiktok_recsys},  entertainment~\cite{amazon_recsys, netflix_recsys, hulu_recsys}, and e-commerce~\cite{amazon_recsys, ebay_recsys, alibaba_recsys}. Modern recommendation systems provide personalized suggestions to enhance user experience through Deep Learning Recommendation Models (DLRM)~\cite{naumov2019deep}.
The growing importance of DLRMs is evident in their widespread deployment by hyperscalers for both training and inference. This translates to a significant portion of AI inference cycles being dedicated to DLRMs~\cite{gupta2020architectural}, while being deployed on a variety of platforms including CPUs(~\cite{jain2023optimizing, nair2024parallelization, gupta2020architectural, mlperf_submissions, jha2024mem}), GPUs(~\cite{gupta2020deeprecsys, ke2022hercules, hsia2020cross, firoozshahian2023mtia, mlperf_submissions}), and accelerators(~\cite{firoozshahian2023mtia, ke2020recnmp, hwang2020centaur, gupta2021recpipe, kal2021space, ke2021near, kwon2019tensordimm}). With the ever-increasing compute and memory requirements of DLRMs, it is increasingly being preferred to execute them on GPUs~\cite{firoozshahian2023mtia, mlperf_submissions} due to their efficient parallel processing capabilities. However, with growing model and dataset sizes, efficient utilization of GPUs for improving the performance of inference applications, as will be shown in this paper, is insufficiently investigated.

\begin{figure}
\centering
\includegraphics[width=0.9\columnwidth]{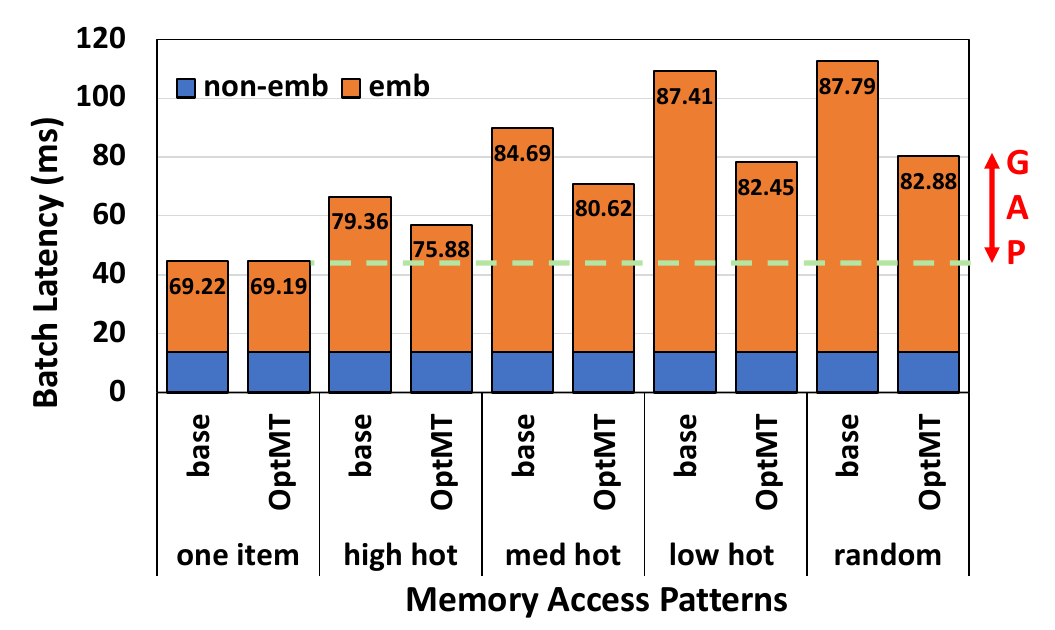}
\caption{Shown is the degradation in inference performance as hotness lowers (working footprint decreases) from left to right. The numbers inside the bars indicate the embedding stage contributions. Here, OptMT provides higher WLP which enhances performance over off-the-shelf PyTorch (base). Yet,  a significant gap continues to exist compared to the fastest loads (one item case). We cite this as the research \emph{gap}.}
\label{fig:intro}
\vspace{-0.2in} 
\end{figure} 

% \textcolor{orange}{Rishabh: I will compress this text. The aim is to connect DLRM application to the platforms used for deployment, and highlight why GPUs are used.}

%meeting real-time service-level agreements (SLAs) for user-facing inference applications remains challenging \textcolor{red}{add one line about WHY this is challenging}. %This necessitates the development of innovative solutions to bridge this performance gap.
%Thus, with their rapid adoption\cite{}, there is a commensurate rise in hardware support with upcoming dedicated ASICs(TPU-v4, MTIA) and design changes in GPUs(Nvidia H100).

DLRMs primarily comprise four stages: embedding, bottom multi-layer perceptron (MLP), feature interaction, and top MLP. The latter three stages are marked as non-embedding stages. Prior works~\cite{jain2023optimizing,nair2024parallelization, gupta2020architectural, gupta2020deeprecsys, hsia2020cross} have shown that the embedding stage is memory intensive (due to frequent and irregular memory accesses) and the non-embedding stages are compute intensive. 

Using the latest off-the-shelf PyTorch-based embedding bag 
CUDA kernels over an A100 GPU, we conduct an extensive characterization study over the latest DLRMs with production datasets. These experiments reiterate that the  embedding stage continues to bottleneck the inference performance as shown in Figure~\ref{fig:intro}. 
%As production datasets exhibit a spectrum of memory access patterns~\cite{gupta2020architectural, gupta2020deeprecsys, jain2023optimizing, sethi2022recshard}, categorizing them based on their degree of hotness, they can range from one item to random. 
GPUs are well-known for their Multi-Threading (MT) or Warp-Level-Parallelism (WLP)\footnote{We use MT and WLP interchangeably in the paper.} support to hide memory latencies with computation~\cite{why_gpu}. Unfortunately, the available parallelism is not enough. We observe up to 3.2$\times$ embedding-only performance slowdown when comparing `random' with the `one item' case (referring to  the fastest case where all embedding accesses point to one row in a table, leading to $\sim$100\% cache hits). This sub-optimal MT is observed in both ready-made packaged and source-compiled PyTorch implementations as they suffer from register pressure (details mentioned in Section~\ref{subsec:characterization}).

To test if increasing the parallelism can address this issue, we synthetically increased the parallelism level to an optimal amount (Optimal MT (OptMT)).\footnote{Note that due to register spilling optimal MT may not be the maximum WLP supported by the GPU. We quantify this effect in Figure~\ref{fig:vary_wlp}.} We did so by lowering the register allocation per warp using available compiler optimizations. Although it did improve the performance (reduction in batch latency) by 53\%, as seen in Figure~\ref{fig:intro}, OptMT is insufficient as a significant performance gap continues to remain between the `one item' and `random' cases. Looking under the hood with detailed profiling (described in Table~\ref{tab:base_profiling}), we observe that both off-the-shelf PyTorch and OptMT implementations underutilize the ``warp issue slots" and ``average HBM read bandwidth", demonstrating that the kernel is memory latency bound.

Previous DLRM-based works have cited the memory-bound issue arising due to embeddings, and have developed heterogeneous platform-based scheduling frameworks~\cite{gupta2020deeprecsys,hsia2020cross,ke2022hercules, ke2022disaggrec}, distributed inference strategies~\cite{lui2021understanding, matam2024quickupdate}, accelerator designs~\cite{firoozshahian2023mtia, ke2020recnmp, hwang2020centaur, gupta2021recpipe, kal2021space, ke2021near, kwon2019tensordimm}, and algorithmic-system designs~\cite{ye2023grace,adnan2021accelerating, kwon2022training, zha2022dreamshard}. However, they are limited to either using out-of-the-box kernels or highly skewed (exhibiting high temporal reuse) datasets, and \emph{none of the prior solutions address the long latency load stalls arising on a GPU platform}. Towards this, we explore easy-to-adopt design solutions by asking the following question:~\emph{Given the high adoption of GPUs by  hyperscalars~\cite{meta_purchasing_gpu} even while being expensive, can we develop cost-friendly software techniques that are both application- and architecture-aware to alleviate the memory bottleneck?} 

Building on our understanding of the unique characteristics of DLRMs, we leverage the features of modern GPUs to make the following contributions: 
%Leveraging the capabilities of modern GPUs, including support for scoreboarding mechanism \textcolor{red}{mention briefly what this is}, large cache capacities (Table~\ref{tab:gpu-cache-capacities}), and L2 cache residency control, we present two novel software designs to overcome the memory challenge: (1) {\em Prefetching to hide the long latency  stalls,} and (2) {\em L2 cache pinning to reduce memory access latency} (Table~\ref{tab:access-latency-gpu-mem}). By taking advantage of the available bandwidth headroom, we carefully tailor prefetching for the embedding bag CUDA kernel, improving embedding lookups up to 97\% and end-to-end inference up to 73\%. Further, by leveraging modern GPUs L2 cache residency control and power-law distribution of memory accesses, we pin top frequency accessed embeddings onto L2, improving embedding lookups by up to 62\%, and end-to-end inference by up to 48\%. Moreover, pinning and prefetching {\em complement} each other, and when combined, improve embedding lookups by up to 103\%, and end-to-end inference by 77\%. Finally, with their synergy, the worst-case performance gap significantly lowers by 163\% over base PyTorch, and 53\% over OptMT.

%Motivated to achieve performant inference, we consider a powerful data-center grade GPU~\cite{a100_gpu} and use the latest advancements of large DLRMs~\cite{??} with high pooling-factor industrial datasets~\cite{dlrm-dataset, jain2023optimizing}, to make 
\begin{itemize}
  \item 
% \anand{If a reviewer asks saying "out-of-the-box is not necessarily known to be efficient on all GPU hardware and some customization is needed", do we have an answer?}

% \textcolor{orange}{Rishabh: out-of-the-box PyTorch running on GPUs has been used by: (1) meta's dlrm benchmarking repository PARAM (2) MLPerf submissions (3) meta's latest DLRM papers I know like their ASPLOS'23 paper (4) The public codebase of embedding bag was last updated 2 months ago -- so it is up to date, and we also use latest PyTorch}

 To the best of our knowledge, ours is the \emph{first work to study the architectural implications of DLRM inference on GPUs} and to point out the~\emph{microarchitectural inefficiencies leading to memory latency bottleneck (our research gap).} Our in-depth \emph{characterization} shows that out-of-the-box PyTorch DLRM implementation has several performance-related inefficiencies. First, out-of-the-box kernel is plagued with long latency load stalls (later described as long scoreboard stalls), leading to a significant performance gap across the spectrum of memory access patterns. Second, the kernel suffers from register pressure leading to limited WLP. The number of hardware registers is simply not enough to support the maximum number of warps allowable. Even if we allow an optimal number of warps by reducing register allocation per warp, there is plenty of scope for reducing latency further. That is, even optimal number of warps is simply {\em not}  enough to hide the long memory latencies. 
  
  % \anand{Not clear what do you mean by "optimal" number in prev sentence. Don't you want as many as possible to hide latency? Would performance worsen beyond a point?}

  % \textcolor{orange}{Rishabh: I point to figure 6 how we pick optimal}

  \item We show that memory latency (not bandwidth) is a major performance-limiting issue for DLRM performance (embedding bag CUDA kernel). On top of optimal WLP, we present two plug-and-play hardware-software co-design based optimizations: (i) by leveraging the bandwidth headroom of modern GPUs, their hardware-supported scoreboarding, and multiple memory resources as buffer stations, we perform {\bf Prefetching} to hide memory latency stalls, and (ii) by taking advantage of known apriori power-law distribution  
  in embedding accesses, we perform {\bf L2 Cache Pinning} to pin the most frequently accessed entries by exploiting the new L2 cache residency control feature on GPUs. Further, we show these two designs can complement each other.

  \item We  evaluate and compare the benefits of the proposed techniques. In isolation, prefetching and L2 pinning improve embedding lookups by up to 97\% and 62\%, and end-to-end inference by up to 73\%, and 48\%, respectively.  Moreover, pinning and prefetching {\em complement} each other, and when combined, improve embedding lookups by up to 103\%, and end-to-end inference by 77\%. Finally, with their synergy, the worst-case performance gap significantly lowers by 163\% over base PyTorch, and 53\% over OptMT. Also, we believe our proposed designs can be generally applied to a wide range of memory-bound kernels.
  %We show that prefetching and pinning can symbiotically integrate while bringing up to a 103\% embedding stage-only and 77\% end-to-end DLRM improvement, thus lowering the worst case performance gap by 163\%.
\end{itemize}

\section{Background}
\label{sec:background}
In this section, we discuss (1) the  architecture of a modern recommendation system, (2) the key microarchitectural features of the latest GPUs, and (3) the related works on improving the DLRM and memory-bound kernel for GPUs.

% NOTE: keep adding references while you write

\subsection{DLRM Inference Using GPUs}
\label{subsec:model_arch}
% TODO: 
% \begin{itemize}
%   \item complete model (use previous ISCA paper image as a placeholder)
%   \item mention about the popular trends
%   \item mention about GPU choice
% \end{itemize}

\begin{figure}
\centering
\includegraphics[width=0.7\columnwidth]{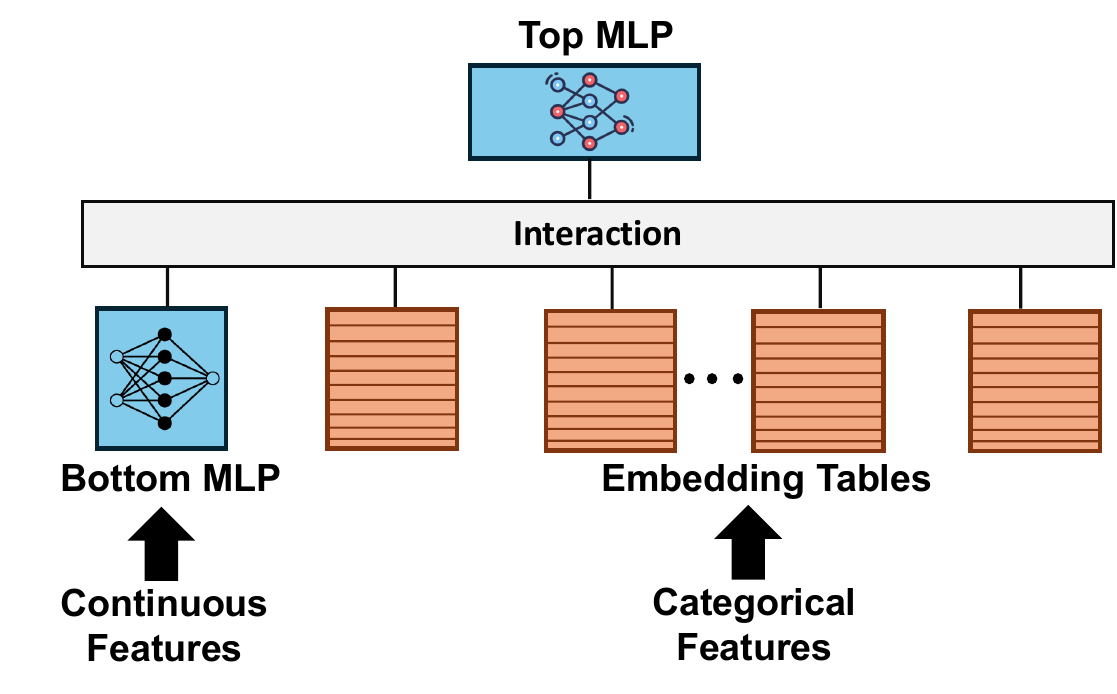}
\caption{A schematic of a DLRM architecture. The continuous features (e.g., age, location) are processed by  Bottom MLP, and categorical features (e.g., movie genre, item ID) by the Embedding Stage. Their outputs are combined in the Feature Interaction Stage, and then fed into the Top MLP, which predicts top-k items with  highest Click Through Rate (CTR).}
\label{fig:dlrm_model_architecture}
\vspace{-0.2in} 
\end{figure}

Many industries use GPUs to execute DLRM inference ~\cite{mlperf_submissions, gupta2020deeprecsys, ke2022hercules, firoozshahian2023mtia}. The primary steps in inference involve (1) a one-time loading of the complete model onto the GPU memory, (2) feeding the input batches (each batch is large and comprises a group of samples) to the GPU, and (3) executing the inference to predict the top-k items for each sample within a batch. 
%Given the real-time and interactive nature of the application, there is a strict SLA requirement~\cite{??} under which an inference must finish. The batch size is carefully chosen to maximize throughput while meeting the SLA target~\cite{??}. 
For large models exceeding the memory capacity of one GPU, multiple GPUs/nodes are used with model and data parallelism~\cite{lai2023adaembed, mudigere2022software}. Regardless of the number of GPUs used, each GPU executes one or more embedding tables serially~\cite{pytorch_dlrm, param_embedding}. 

Figure~\ref{fig:dlrm_model_architecture} shows a simplified diagram of a typical DLRM ~\cite{gupta2020architectural, naumov2019deep}. Each sample comes with continuous (e.g., age, location) and categorical features (e.g., movie genre, item id). Former are fed to the Bottom MLP stage while latter are fed to the Embedding Stage. The feature interaction stage merges (concatenation/dot product) the outputs of the previous two stages, and feeds it to the Top MLP stage, generating the top-k items with the highest predicted CTR (Click-Through Rate). Several past works highlight the embedding stage to be memory intensive (~\cite{kal2021space, ke2020recnmp, jain2023optimizing, nair2024parallelization, ye2023grace}) and non-embedding stages to be compute intensive~\cite{gupta2020deeprecsys}.

\subsection{Key Properties of GPU Microarchitecture}
\label{subsec:gpu_properties}
\begin{figure}
\centering
\includegraphics[width=0.8\columnwidth]{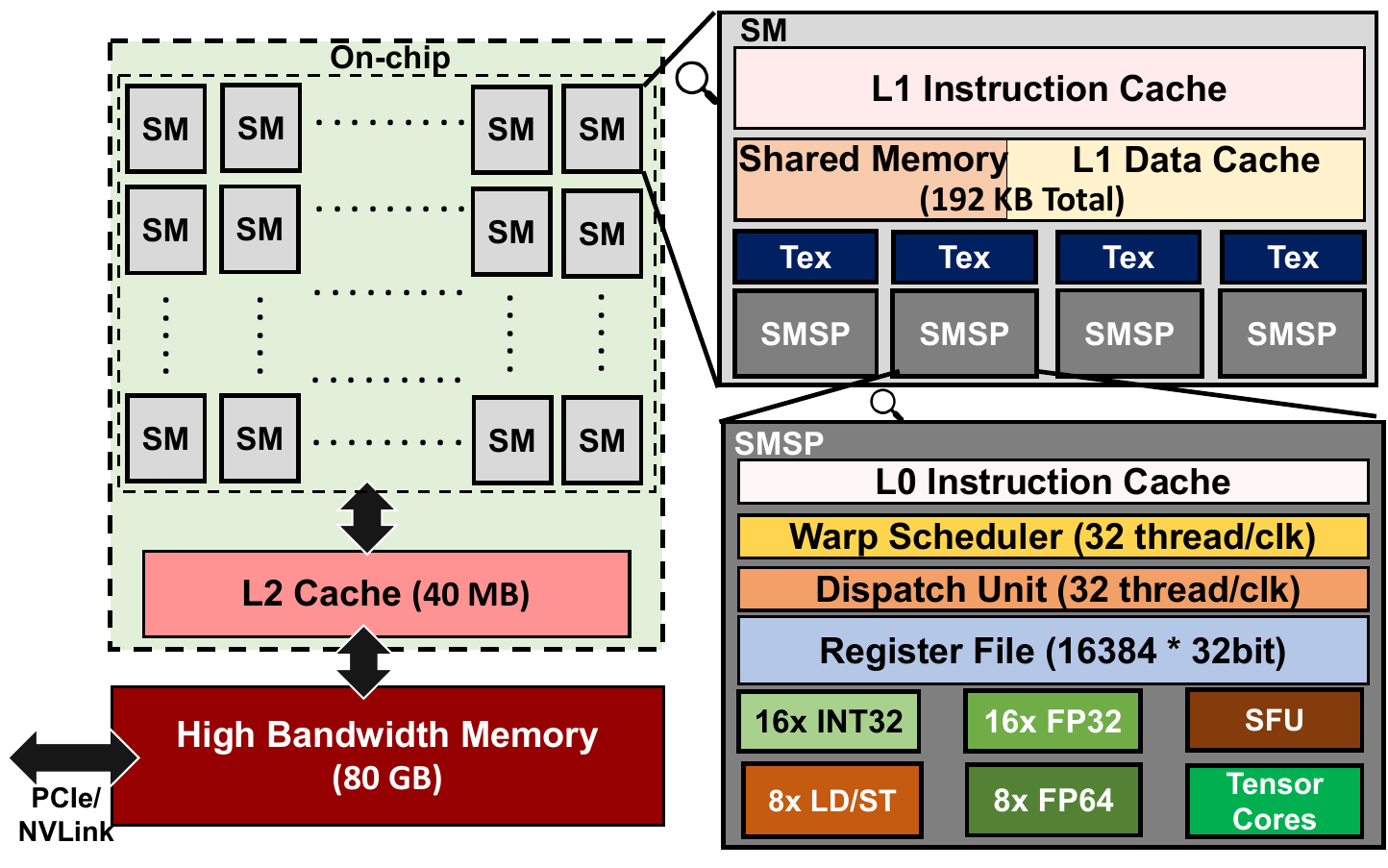}
\caption{Simplified Nvidia A100 GPU organization.}
\label{fig:gpu_organization}
\vspace{-0.2in} 
\end{figure}

GPUs, also known as throughput processing engines, contain a hierarchical array of compute cores (CUDA cores). Figure~\ref{fig:gpu_organization} shows a simplified diagram of the GPU organization. Modern GPUs (Nvidia based) contain 100s of SMs (Streaming Multiprocessors)~\cite{a100_gpu, h100_gpu}, and each SM offers 4 SMSP (Streaming Multiprocessor Sub-Partition). Each SMSP is associated with a warp scheduler and provides the capability to issue one eligible warp every cycle while maintaining a queue of resident warps, thus facilitating WLP or MT. Further, a scoreboarding mechanism~\cite{gpgpu-sim_scoreboard} is adopted in the core pipeline to promote instruction level parallelism (ILP). The memory organization of each SM consists of: (1) a large register file storing the context of all resident warps, allowing zero overhead warp switching and (2) a private cache shared among all residing warps. All SMs share an L2 cache and an off-chip memory. Additionally, Nvidia GPUs (Ampere generation and onwards) provide a unique programmer-controlled L2 access management for setting-aside a region for persisting accesses~\cite{l2_residency}. Section~\ref{subsec:design_pinning} guides on how we take advantage of this feature for a performant embedding stage execution.

The efficiency of DLRM execution is heavily influenced by the unique properties of GPUs. All the stages have parallel implementations to reap benefits of the massive number of CUDA cores. Also, the high bandwidth memory (HBM) helps in meeting the heavy off-chip access requirements for the embedding stage. Additionally, a modest on-chip cache hierarchy helps in capturing the locality in memory accesses. Table~\ref{tab:access-latency-gpu-mem} shows the access latencies~\cite{luo2024benchmarking} for different access locations. Note that fetching data is more costlier compared to CPUs. Table~\ref{tab:gpu-cache-capacities} shows the L1 and L2 cache capacities in server-grade and powerful GPUs. Note that the cache sizes are much larger in the latest GPUs. For example, (1) A100 offers 1.5x and $\sim$7x large sizes over V100 for L1 and L2, respectively, and (2) RTX 4090 offers $\sim$12x large L2 cache over RTX 3090 Ti. 

\begin{table}[!ht]
\caption{Access Latencies for various levels of the A100 GPU memory hierarchy based on~\cite{luo2024benchmarking}.}
    \centering
    %\footnotesize
    \scriptsize
    \setlength{\tabcolsep}{3pt}
    \begin{tabular}{|c|c|}
    \hline
        \textbf{Access Location} & \textbf{Access Latency (cycles)} \\ \hline
        Register & 1 \\ \hline
        Shared Memory & 29 \\ \hline
        L1D cache & 37.9 \\ \hline
        L2 cache & 261.5 \\ \hline
        Global Memory & 466.3 \\ \hline
    \end{tabular}
    
    \label{tab:access-latency-gpu-mem}
\vspace{-4mm}
\end{table}

\begin{table}[!ht]
\caption{Cache capacities for server-grade GPUs.}
    \centering
    %\footnotesize
    \scriptsize
    \setlength{\tabcolsep}{3pt}
    \begin{tabular}{|c|c|c|}
    \hline
        \textbf{Device} & \textbf{LLC cache size (MB)} & \textbf{L1 cache size (KB)} \\ \hline
        %M1300X & 256 & 16 \\ \hline
        A100 & 40 & 192 \\ \hline
        H100 & 50 & 256 \\ \hline
        L40 & 96 & 128 \\ \hline
        RTX4090 & 72 & 128 \\ \hline
    \end{tabular}
    
    \label{tab:gpu-cache-capacities}
\vspace{-6mm}
\end{table}

% In the newer generation of GPUs, 
% Increase in: CUDA cores and SMs
% Connecting with the parallel nature of Embedding stage, GPu is a popular platform, cite several works. 
% TODO
% \begin{itemize}
%   \item one line description of the parallel cores and SMs, and features HBM. For example, give hopper GPU. 
%   \item cuda core pipeline allowing independent instructions to be launched.
%   \item trend of rising cache capacities
%   \item L2 cache residency control
%   \item summarize the CUDA kernel properties of EmbeddingBag Kernel: kernel launch configuration, cuda registers, occupancy, 
% \end{itemize}

\subsection{Related Works}
\label{subsec:related_works}

\textbf{DLRM Optimizations:}
%DLRMs are popularly used in a variety of applications ~\cite{cheng2016wide, naumov2019deep, zhou2018deep, zhou2019deep} and has spurred significant interest from researchers~\cite{mudigere2022software, gupta2020architectural, gupta2020deeprecsys, gupta2021recpipe, sethi2022recshard, hsia2020cross, ke2022hercules}. 
Prior works  
~\cite{gupta2020deeprecsys,hsia2020cross,ke2022hercules, ke2022disaggrec} have looked into scheduling frameworks and heterogeneous platforms for inference serving and ~\cite{lui2021understanding} discusses the system design for effective distributed inference.
~\cite{ye2023grace,adnan2021accelerating, kwon2022training} considers highly skewed (exhibiting high temporal reuse) dataset cases and proposes algorithmic and system designs for effectively using GPU's main memory. 
However, none of the prior works have studied the microarchitectural implications of DLRM inference on a GPU.
Our work improves the embedding table performance via on-chip optimizations for a diverse set of access patterns(Figure~\ref{fig:coverage_study}), making it orthogonal, and thus it can be combined the prior works. 
% \todo{TODO Mention accelerator-based works while we optimize of existing hardware} 

\textbf{Accelerator designs:} ~\cite{firoozshahian2023mtia, ke2020recnmp, hwang2020centaur, gupta2021recpipe, kal2021space, ke2021near, kwon2019tensordimm} have proposed targeted custom solutions for MLP and embedding stages in DLRMs. However, these proposals require substantial time and effort to commercialize, making it difficult to adopt with the fast changing model parameters. With GPUs being widely adopted~\cite{meta_purchasing_gpu}, our plug-and-play solutions can be instantly leveraged (Section~\ref{sec:optimizations}).

\textbf{Scheduling and Virtualization:} 
Several warp and CTA scheduling works have been proposed in the past~\cite{jog2013owl, kayiran2013neither, sethia2015mascar} to improve GPU performance by hiding memory latencies effectively. Most of these works 
focused on improving cache and memory contention or finding optimal thread-level parallelism. We show that 
even on the recent A100 GPUs, latency hiding capability is limited due to the limited register file (Section~\ref{subsec:characterization}). Register-file virtualization techniques~\cite{jeon2015gpu, voitsechov2018software, oh2018finereg} have been proposed in the past to address the issues related to limited register file. However, they are implemented in GPU hardware and often have non-trivial overheads. In contrast, we provide a complementary software-only solution (prefetching and pinning) that is aware of both GPU application and underlying hardware (Section~\ref{sec:optimizations}). 

\textbf{Prefetching on GPUs:} 
Given that GPU memory bandwidth is limited, data prefetching needs to be done carefully to result into any performance benefits. Prior works on GPU prefetching~\cite{jog2013orchestrated, sethia2013apogee, oh2018adaptive, wu2011pacman} consider this issue and show performance improvements. However, to the best of our knowledge, there is no prior work that considers software prefetching in GPUs that is particularly tailored for emerging applications such as 
 DLRM (Section~\ref{subsec:design_prefetching}). 

\textbf{L2 Cache Management:} Recently, with Nvidia's Ampere architecture and onwards~\cite{a100_gpu, h100_gpu}, the GPUs feature a CUDA/PTX-based programmer control for L2 cache management~\cite{l2_residency}. 
%Leveraging L2 pinning is relatively new. 
~\cite{fu2023autoscratch, adufu2023l2, adufu2023optimizing} uses the L2 cache control for improving GEMM, LSTM, fully-connected and convolution-based kernels. In contrast, our paper proposes to apply this feature for embedding stage by pinning the most frequently accessed embeddings (Section~\ref{subsec:design_pinning}).

% Highlight the popularity of DLRMs and various optimizations have been proposed.

% Contrast with out contribution: 
% * No prior work has looked into MT of DLRMs on GPUs
% * No prior work has looked into prefetching for DLRMs
% * No prior work has looked into L2 pinning for DLRMs
% TODO: Connect with the Novelty section, and compare with the other works in contrasting manner. Clear aim: (1) establish that none of the prior works have looked into the microarchitectural implications of embedding stage on a GPU (2) many works have cited memory bandwidth as a problem: L1/L2 caches would offer in TBps, but off-chip BW is not well utilized. (3) other works which improve inference/training by smart scheduling or algorithm, our work can be combined with them -- qualitiative answer, and also mention that our work focuses on optimizing on-chip computation (1GPU). 

%%%%%
\section{Dissecting Embedding Bag execution on a GPU}
\label{sec:motivation}
%Section~\ref{sec:background} presented a high-level overview of the DLRM architecture and discussed the key properties of the latest GPUs. 
For better understanding the inference behavior on GPUs, this section discusses: (1) the parallel implementation, work partitioning, and mapping of the embedding stage on CUDA threads; (2) a quantitative study of memory access patterns used in production deployments; and (3) the architectural implications of off-the-shelf and optimal-MT PyTorch-based DLRM inference on GPUs, catering to a variety of memory accesses. Finally, we conclude that memory latency continues to remain a challenge, and motivate towards optimizations addressing this issue to achieve better performance.

%\vspace{-5mm}
\subsection{Parallel Implementation of the Embedding Bag Operator}
\label{subsec:work_partitioning}
The embedding stage of DLRM involves numerous parameters, and understanding how each one affects the performance is crucial. To illustrate this, Algorithm~\ref{algo:algorithmI} highlights the high-level working of the embedding stage. Arriving queries create ``batches", where each batch is expected to meet the SLA target. For each table, the batch contains a batch size (BS) number of samples, and each sample involves a pooling factor (or lookups per sample) amount of work over the embedding vectors of length equal to the embedding dimension (ED). At the core, each sample does a gather (load) and reduce (accumulation) operation. The amount of data processed in each table can be calculated as (BS) $\times$ (average lookups per sample) $\times$ (ED) $\times$ (precision). For example, for our chosen configuration (described in Section~\ref{sec:methodology}), the amount of data processed per table is 2048 $\times$ 150 $\times$ 128 $\times$ 4B = 150 MB. Consequently, the complete embedding stage processes 37.5 GB of data. With the inference running on a GPU, the gather-reduce operations are executed in a Single-Instruction-Multiple-Threads (SIMT) manner, thus exploiting parallelism.

\begin{algorithm}
%\scriptsize
%\small
\footnotesize
\begin{flushleft}
\For{v in 0 ... num\_batches}
{
  \For{w in 0 ... num\_tables}
  {
    \For{x in 0 ... batch\_size}
    {
      \For{y in 0 ... lookups\_per\_sample}
      {
        \vspace{1mm}
        \textbf{SIMT load} accm on register;
        %\vspace{1mm}
        
        \For{z in 0 ... embedding\_dim}
        {
          \textcolor{red}{\textbf{SIMT load} row\_block on register;}
          
          \textbf{SIMT add} accm, row\_block;
        }
        \vspace{1mm}
        \textbf{SIMT store} accm to memory;
      }
    }
  }
}
\end{flushleft}
\caption{\small Simplified memory access loop for the embedding stage on GPU.}
\label{algo:algorithmI}
\end{algorithm}

\begin{figure}
\centering
\includegraphics[width=1\columnwidth]{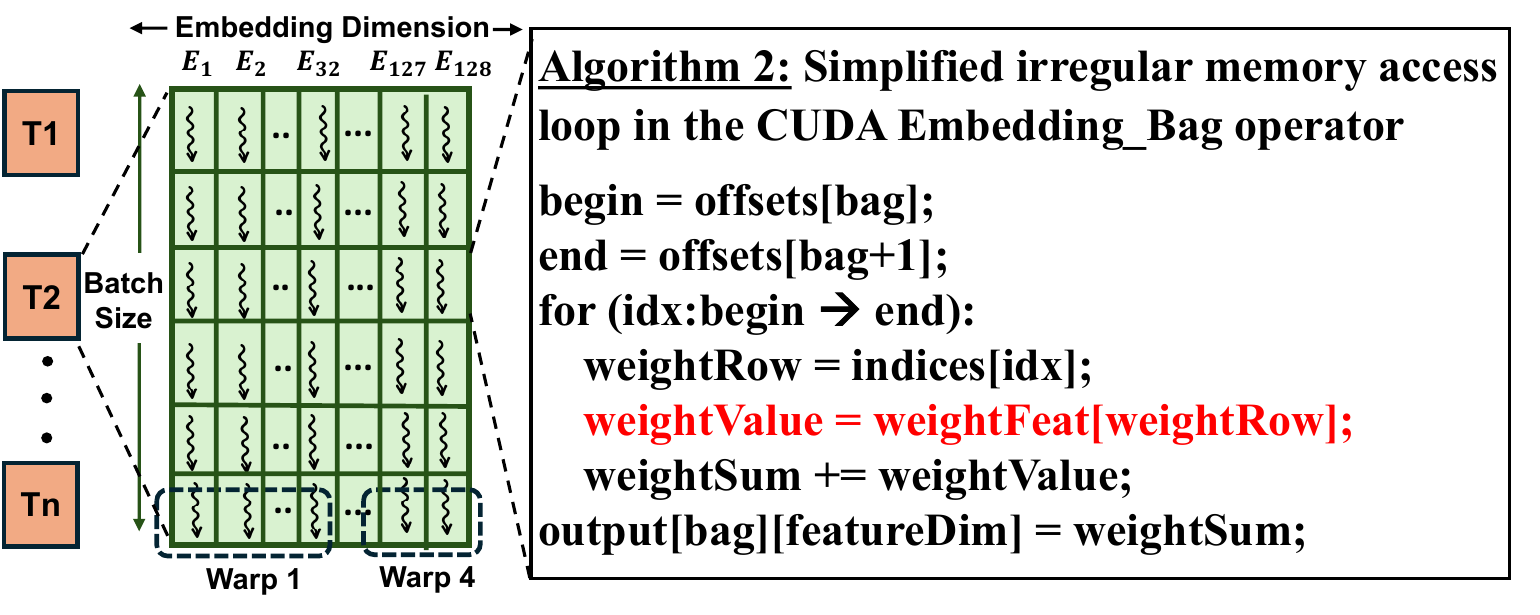}
\caption{Parallel implementation of embedding stage by work partitioning across CUDA threads. Here, 1000s of CUDA threads independently work on one output matrix element.}
\label{fig:work_partitioning}
\vspace{-5mm} 
\end{figure}

To better understand the incorporated parallelism, Figure~\ref{fig:work_partitioning} breaks down the embedding stage execution into three parts. First, for any number of embedding tables to be completed by one GPU, they are processed sequentially. Second, the embedding bag operator is used to process a table (using PyTorch's backend CUDA kernel \textit{"EmbeddingBag\_updateOutputKernel\_sum\_mean"}~\cite{emb_bag_cuda_kernel}), which generates an output matrix of dimension (BS) $\times$ (ED). Intuitively, we can visualize that, within a batch, each sample and each embedding element is independent of the other. Thus, a CUDA thread works on each embedding element. In this off-the-shelf kernel, we note a static execution launch configuration with a grid size of (1024,1,1) and a block size of (32,8,1). This results in a large number of CUDA threads, and thus, fully uses all the SMs provided in the latest GPUs (e.g., the 108 SMs in A100). Warps are automatically formed in the GPU by combining adjacent CUDA threads. For example, with an embedding dimension of 128, 4 warps are formed to process a sample. Third, Algorithm 2 (Figure~\ref{fig:work_partitioning})  highlights the work within a thread which encapsulates a ``number of lookups" amount of gather-reduce operations (\textcolor{black}{thus each thread partially completes the two innermost loops in Algorithm~\ref{algo:algorithmI}}). Fundamentally, the gather is similar to a pointer-chasing operation as we access a series of arrays to complete it (the offset array, followed by the indices array, followed by the embedding table). Thus, this operation results in the irregular loads which have a significant impact on (and causes variation in) the performance of the embedding stage (Figure~\ref{fig:intro}). 
%\todo{Describe coalescing of loads and no branch divergence}

% \begin{figure}
% \centering
% \includegraphics[width=0.7\columnwidth]{figures/coverage_study.pdf}
% \caption{Coverage study for different memory access patterns: it shows the \% of total accesses (y axis) that are covered by the \% of unique accesses (x axis).}
% \label{fig:coverage_study}
% \vspace{-5mm} 
% \end{figure}

\subsection{Quantitative Study of the Memory Access Patterns}
\vspace{-1mm}
\label{subsec:memory_access_patterns}
Embedding accesses in DLRMs follow a ``power-law" distribution where a small portion of embedding table entries services a large fraction of accesses~\cite{gupta2020deeprecsys, gupta2020architectural, jain2023optimizing}. In our study, we use the datasets from a recent work~\cite{jain2023optimizing},  which extracts homogeneous datasets using Meta's production traces~\cite{dlrm-dataset}.  

% \setlength{\columnsep}{4mm}
% \setlength{\intextsep}{4mm}
% \begin{wrapfigure}{R}{0.25\textwidth}
% \centering
%     \includegraphics[width=0.25\textwidth]{figures/coverage_study_new.pdf}
%     %\vspace{-7mm}
%     \caption{Coverage study for different memory access patterns: it shows the \% of total accesses (y axis) that are covered by the \% of unique accesses (x axis).}
%     \label{fig:coverage_study}
%     \vspace{-1mm}
% \end{wrapfigure}

\begin{figure}
\centering
\includegraphics[width=0.75\columnwidth]{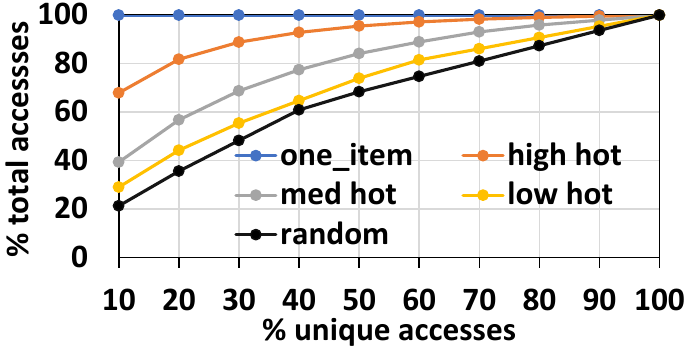}
\caption{Coverage study for different memory access patterns: it shows the \% of total accesses (y axis) that are covered by the \% of unique accesses (x axis).}
\label{fig:coverage_study}
\vspace{-5mm} 
\end{figure}

Building on prior works~\cite{gupta2020architectural, gupta2020deeprecsys, jain2023optimizing, sethi2022recshard}, we investigate various memory access patterns encountered in real-world industrial settings and categorize them based on their degree of "hotness". To understand hotness, we define two metrics that classify memory access patterns within datasets: unique access \% and coverage study. For a given table, unique access \% represents the proportion of distinct accesses compared to the total number of accesses. Essentially, it measures the variety across  memory locations accessed within the table. Thus, considering a total of R accesses (number of rows in a table) and U unique accesses, the unique access \% is calculated as U$\times$100/R. Table~\ref{tab:unique_accesses_per_dataset} shows each dataset's unique access \%. Note that one item and random are ``synthetic" datasets; the former corresponding to the case where all indices match and point to the same entry in a table, whereas the latter means all indices are uniformly distributed within a range of [0, R). Thus, unique accesses range between 0 to 100\%, being lowest for one item and highest for random.

\begin{table}[!ht]
 \caption{Unique access \% in each dataset.}
    \centering
    %\footnotesize
    \scriptsize
    \setlength{\tabcolsep}{6pt}
    \begin{tabular}{|c|c|c|c|c|c|}
    \hline
        \textbf{Datasets} &\textbf{one item}&\textbf{high hot}&\textbf{med hot}&\textbf{low hot}&\textbf{random}\\ \hline
       \textbf{unique access\%} & 0.0002 & 4.05 & 20.50 & 46.21 & 63.21 \\ \hline
    \end{tabular}
   
    \label{tab:unique_accesses_per_dataset}
\vspace{-2mm}
\end{table}

Further, the distribution of the unique accesses influences the actual memory access pattern. Figure~\ref{fig:coverage_study} shows the coverage study by noting how much total accesses get covered by a given number of \% unique access.
%(being $\geq$ 1). 
For example, in the one item case, one embedding covers all 100\% accesses (making the trend across x-axis uniform), whereas, in the high hot case, 10\% of the total unique items are sufficient to capture 68\% of the total accesses. Regardless of the hotness, it is important to note that the total memory access count remains the same in each of these datasets. Thus, it is fair to compare the performance of different datasets while ensuring the same amount of observed loads. Using these two metrics, it can be noted that, for a given table, the hotness decreases from 'one item' to 'random',  causing an increase in the working set size and total number of irregular loads. 
\vspace{-1mm}
\subsection{Architectural Implications of Embedding Bag on a GPU}
\label{subsec:characterization} 
Previous subsections have highlighted the amount of parallelism offered in the embedding bag operator and how it leverages the GPU for execution. Modern GPUs~\cite{A100, H100} provide larger caches and HBM capacity,  which directly help embeddings data reuse behavior and bandwidth needs. In this spirit, various previous works~\cite{gupta2020deeprecsys, ke2022hercules, lai2023adaembed, sethi2022recshard, lin2022building, hsia2020cross} have used GPUs for DLRM inference and training. However, to our knowledge, no prior work has conducted a detailed profiling to study the 'microarchitecture behavior' of DLRM execution on GPUs. Specifically, given that DLRMs are generally memory bound~\cite{gupta2020deeprecsys, jain2023optimizing, gupta2020architectural}, it is important to thoroughly verify whether the primary application kernel is effectively utilizing the GPU's resources. With this motivation, we carefully investigate the embedding bag kernel Figure~\ref{fig:work_partitioning}~\cite{emb_bag_cuda_kernel} using Nvidia's Nsight Compute Tool (NCU) on an A100 80 GB GPU. Since various memory access patterns affect the memory-bound behavior, we evaluate multiple datasets (Table~\ref{tab:unique_accesses_per_dataset}).  

\begin{table}[!ht]
\caption{Microarchitectural characterization of Base PyTorch on various datasets. With 74 registers allocated to each CUDA thread, the WLP is limited due to the register pressure.}
\vspace{-2mm}    
    \centering
    %\footnotesize
    \scriptsize
    \setlength{\tabcolsep}{6pt}
    
    %\begin{tabular}{|c|c|c|c|c|c|}
    \begin{tabular}{|p{3.8cm}|p{0.4cm}|p{0.4cm}|p{0.4cm}|p{0.4cm}|p{0.6cm}|}
    \hline

    \textbf{NCU metrics/datasets}                           & \textbf{one item} & \textbf{high hot} & \textbf{med hot} & \textbf{low hot} & \textbf{random} \\ \hline
    \textbf{Kernel time (us)}                    & 138 & 237 & 341 & 428 & 442 \\ \hline 
    \textbf{\#load insts (M)}                 & 2.47     & 2.47     & 2.47     & 2.47     & 2.47     \\ \hline
    \textbf{SM Throughput \%}             & 71.45    & 41.27    & 26.65    & 21.23    & 20.42    \\ \hline
    \textbf{warp cycles per executed inst}    & 7.06     & 11.7     & 17.56    & 21.94    & 22.86    \\ \hline
    \textbf{long scoreboard stall   (cycles)} & 1        & 7.2      & 13.1     & 17.7     & 18.6     \\ \hline
    \textbf{issued warp per scheduler per cycle}        & 0.77     & 0.47     & 0.31     & 0.25     & 0.24     \\ \hline
    \textbf{Global L1\$ hit rate \%}          & 98.7     & 42.74    & 30.11    & 20.36    & 19       \\ \hline
    \textbf{L2\$ hit rate \%}                 & 99.46    & 93.96    & 59.5     & 18.71    & 7.7      \\ \hline
    \textbf{Device Memory size read(MB)}              & 0        & 4.87     & 45.96    & 122      & 144.57   \\ \hline
    \makecell{\textbf{Avg HBM Read BW(GBps)} } & $\sim$0           & 20.8              & 135              & 286.5            & 329.5 \\\hline
    \textbf{Avg HBM Read BW Utilization (\%)} & $\sim$0           & 1.04              & 6.75              & 14.33            & 16.5 \\\hline

    \end{tabular}
    
    \label{tab:base_profiling}
\vspace{-0mm}
\end{table}

Table~\ref{tab:base_profiling} describes the off-the-shelf PyTorch characterization using various NCU metrics. Recall that Figure~\ref{fig:intro} highlighted that random performs 3.2$\times$ slower than the fastest one item case, even though both observe the same number of loads. This is because the SM or compute throughput is heavily impacted by random accesses. With the decrease in hotness (one item to random), the data reuse gets reduced~\cite{jain2023optimizing}, causing an increase in the warp cycles per executed instruction. As each CUDA thread performs a pooling factor amount of gather-reduce operations, a load-use dependency arises. We look into the breakdown of warp cycles and inspect the long scoreboard stall cycles to exactly capture these dependency stalls. For all datasets except one item, the warp cycles are mainly constituted from the long  scoreboard stalls. The absolute stall cycles are impacted by the amount of data captured by the caches. Note that both warp cycles per executed instruction and long scoreboard stalls are {\em averaged} over all executed instructions, and so they cannot be directly compared to  the memory latency values in Table~\ref{tab:access-latency-gpu-mem}. As one item dataset has a minimal working set (512B), it experiences much lower stalls due to best cache locality. However, both L1 and L2 cache hit rates significantly drop as the hotness lowers, increasing the amount of data read from the device memory. Therefore, the average bandwidth demand is significantly higher towards the random dataset, reaching up to 329.5GBps. Further, the peak read bandwidth (measured using Nvidia Nsight Systems~\cite{nvidia_nsight_systems}) achieved is 510GBps for the random case. However, this observed bandwidth is small compared to the theoretical peak bandwidth of HBM (2TBps). This significant disparity makes us suggest that the Embedding Bag operator is a memory latency-bound kernel. 

The Nvidia A100 GPU is based on compute capability 8~\cite{A100}, meaning that one SM houses up to a maximum of 64 resident warps. These resident warps enable WLP primarily helping in hiding any kinds of stalls. 
%For example, if the current warp encounters a dependent instruction with unresolved operands, the warp scheduler switches the execution to a resident ready-to-execute warp with zero overheads~\cite{??}.
We observe that the PyTorch CUDA kernel~\cite{emb_bag_cuda_kernel} uses a high number of registers (74), and thus suffers from the register pressure, leading to a low theoretical occupancy of 37.5\% (or 24 resident warps per SM). Figure~\ref{fig:gpu_organization} indicated that one SM contains a total of 4 warp schedulers, meaning that each scheduler gets to work with only 6 warps, even though the hardware supports a maximum of 16 warps. The metric ``issued warp per scheduler per cycle" (also called as ``issue slot utilization") captures the number of issued warps every cycle, which is a function of both WLP and warp cycles per issued instruction. Similar to the SM throughput, it decreases as the hotness lowers. Thus, even though the CUDA kernel encounters significant long scoreboard stalls, the application lacks in providing effective WLP, limiting the hardware's capability to hide these stalls.

\begin{figure}
\centering
\includegraphics[width=0.8\columnwidth]{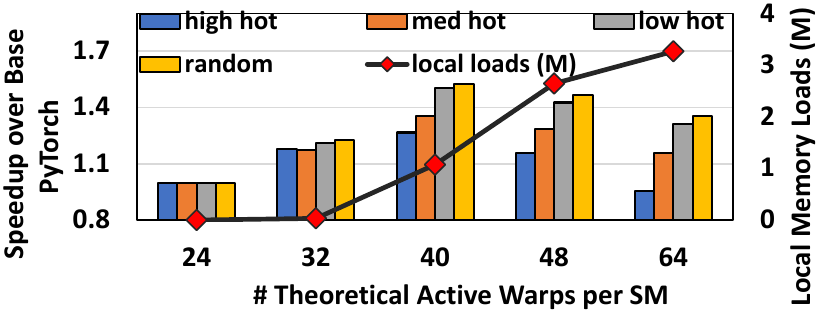}
\caption{Synthetically varying the number of registers allocated to improve WLP. The primary y-axis is speedup over off-the-shelf PyTorch, and the secondary y-axis is the register spilling penalty based on extra local memory loads (in millions). OptMT refers to the highest speedup at 40 warps.}
\label{fig:vary_wlp}
\vspace{-2mm} 
\end{figure}

Since higher WLP could potentially better mitigate the memory latency, we force the compiler to strategically {\em limit} the allocated registers during compilation, resulting in more warps to be resident in one SM, eventually improving the WLP. To achieve this, we compile PyTorch with ``-maxrregcount maxreg"~\cite{nvcc_maxrregcount} flag, where the ``maxreg" amount of allocated registers is enforced by the compiler. However, with lower registers in use, now register spilling occurs. The compiler spills the registers to local memory which in turn penalizes the performance. By varying the number of registers, we can sweep through different WLP configurations as seen in Figure~\ref{fig:vary_wlp}. Here, we capture the performance improvement over different datasets. Higher WLP helps in gaining performance with maximum gain at 40 resident warps (denoted as OptMT). Also, higher improvements are seen for low hot and random cases as they require more latency hiding. Further, even though 48 and 64 resident warps provide better WLP, the performance drops due to an increase in register spilling. The impact of register spilling is measured in terms of local memory loads. In the baseline PyTorch (24 warps per SM), all loads/store accesses go to the global memory and none to the local memory, meaning that all the embedding accesses are served from the global memory. However, with an increase in WLP, the register spilling increases, causing an increase in the local memory loads which hurts performance. Thus, there is a clear tradeoff between WLP and the spilling penalty. For instance, in the high hot case, 64 resident warps per SM underperform (compared to the baseline) as the spilling penalty overshadows the potential benefits from multi-threading. 

\begin{table}[!ht]
\caption{Microarchitectural characterization of Optimal-Multithreading (OptMT) PyTorch on various datasets. With 42 registers allocated to each CUDA thread, the register  pressure lowers and the WLP significantly improves. Still, a performance gap exists between the fastest and slowest loads.}
\vspace{-2mm}
    \centering
    %\footnotesize
    \scriptsize
    \setlength{\tabcolsep}{6pt}
    
    %\begin{tabular}{|c|c|c|c|c|c|}
    \begin{tabular}{|p{3.8cm}|p{0.4cm}|p{0.4cm}|p{0.4cm}|p{0.4cm}|p{0.6cm}|}
    \hline

    \textbf{NCU metrics/datasets}                           & \textbf{one item} & \textbf{high hot} & \textbf{med hot} & \textbf{low hot} & \textbf{random} \\ \hline
    \textbf{Kernel time (us)}                    & 135 & 189 & 250 & 282 & 290 \\ \hline 
    \textbf{\#load insts (M)}                 & 3.54    & 3.54     & 3.54     & 3.54     & 3.54     \\ \hline
    \textbf{SM Throughput \%}              & 71.89   & 54.93    & 39.3     & 34.72    & 33.84     \\ \hline
    \textbf{warp cycles per executed inst}    & 10.61   & 15.2     & 20.93    & 24.74    & 25.44     \\ \hline
    \textbf{long scoreboard stall   (cycles)} & 1.33    & 8.6      & 15.3     & 19.6     & 20.4    \\ \hline
    \textbf{issued warp per scheduler per cycle}        & 0.79    & 0.59     & 0.42     & 0.36     & 0.35     \\ \hline
    \textbf{Global L1\$ hit rate\%}          & 98.7    & 37       & 27.2     & 19.85    & 19        \\ \hline
    \textbf{L2\$ hit rate \%}                 & 85.36   & 92.3     & 56.51    & 16.48    & 7.1     \\ \hline
    \textbf{Device Memory size read(MB)}              & 0.3     & 7.5      & 54.1     & 131.9    & 151    \\ \hline
    \makecell{\textbf{Avg HBM Read BW(GBps)} }          & 2.57    & 43       & 226.5    & 485.4    & 547.5   \\
    \hline
    \textbf{Avg HBM Read BW Utilization (\%)} & $\sim$0           & 2.2             & 11.3              & 24.3           & 27.4 \\\hline

    \end{tabular}
    
    \label{tab:OptMT_profiling}
\vspace{0mm}
\end{table}  

Table~\ref{tab:OptMT_profiling} describes the microarchitectural characterization for OptMT. For one item, the performance matches the off-the-shelf PyTorch. For the remaining datasets, the performance significantly improves with the rise in SM throughput. The metric ``warp cycles per executed inst" is slightly higher compared to the baseline for two reasons: (1) with the increase in more resident warps, there exist times when multiple warps are ready and not selected, thus causing an increase in the ``Stall Not Selected" stalls; (2) with the increase in warp switching and local memory accesses, more cache thrashing occurs for global accesses as visible with a slight decrease in cache hit rates, leading to a slight increase in the ``long scoreboard stalls". Based on the latter reason, the total reads from device memory also slightly increases. Finally, the HBM read bandwidth increases to meet the higher demand from WLP.
% However, even though performance has improved compared to the baseline, the ``issued warp per scheduler" per cycle for the random dataset is still far away from the one\_item case. Thus, even with higher optimal WLP, memory latency problem still persists.
While our approach using limited register allocation demonstrates performance gains compared to the baseline, the ``issue slot utilization" for the lower hotness cases remains significantly less than the one-item case. This observation  suggests that {\em even with the enhanced warp-level parallelism (WLP), memory latency continues to be a bottleneck.} 

\subsection{GPU-specific Key Microarchitectural Insights} 
\label{subsec:motivation} 
The following key insights emerge from the preceding microarchitectural characterization study on an A100 GPU:  
\begin{itemize} 
  \item Due to the variations in memory access patterns, a significant performance gap could exist across the spectrum of memory access patterns. In the baseline (or off-the-shelf) PyTorch, this gap is visible and arises from the long  scoreboard stalls due to high cache misses.  
  \item Although GPUs are equipped with high multi-threading support, the off-the-shelf CUDA kernel suffers from register pressure and fails to provide enough WLP.  
  \item The WLP can be increased by forcing the compiler to lower the allocated registers. However, it comes at the cost  of a performance penalty from register spilling. Peak performance gain is seen with 40 resident warps (marked as OptMT).
  \item OptMT lowers batch latency by up to 53\% over the baseline with visible benefits in the SM throughput. Yet, the ``issue slot utilization" continues to see a high gap between the fastest and slowest loads. 
  \item The average read memory bandwidth increases with OptMT, yet it remains small compared to the peak of HBM. Thus, we can conclude that {\em the embedding bag operator is memory latency bound on the GPU.} 
\end{itemize}

Thus, the best WLP contained within an application is insufficient in fully hiding the long latency loads. In the following section, we discuss two complementary techniques to minimize the memory latency further.

\section{Optimizations to Improve the Warp Scheduler Issue Slot Utilization}
\label{sec:optimizations} 
For bridging the gap in the issue slot utilization across the ends of the memory access spectrum, in this section, we first describe the limitations of existing hardware and software approaches. We then propose an application-driven software prefetching strategy and discuss how it can take advantage of various memory resources as buffering stations. Following that, we propose an L2 pinning strategy to overcome the potential shortcomings of prefetching. Finally, we describe how prefetching and pinning can work synergistically.
\vspace{-2mm}
\subsection{Limitations of Off-the-Shelf Solutions}
To deal with memory bound kernels, GPUs employ: (1) WLP and zero-overhead context switching to keep executing useful work even when a warp stalls; (2) large cache line sizes (128 bytes) which help in exploiting spatial locality; (3) a scoreboarding  mechanism which helps to issue and execute consecutive instructions without stalling the pipeline until a dependent instruction is reached; and (4) scratch-pad memory and L2 cache with explicit programmer control for intelligent data placement and reuse. While the off-the-shelf CUDA kernel takes advantage of (1), (2) and (3), it still lacks in performance, as previously highlighted in  Section~\ref{subsec:motivation}.

To take advantage of (3) and (4), an optimizing compiler or application-specific software development can help. An optimizing compiler would try to come up with a valid reordering of instructions to improve performance. For instance, compiler could hoist independent instructions between a load and use dependency and thus promote scoreboarding to minimize pipeline stalls. In this direction, {\em loop unrolling} could be useful as it broadens the scope of finding independent instructions using later iterations. However, when testing the optimal compiler (in O3 level), we note that inserting ``\#pragma unroll" does not have any positive impact on performance because of the runtime-dependent loop bounds (Figure~\ref{fig:work_partitioning}). Furthermore, the compiler cannot directly manage the scratch pad or L2. We believe these challenges open the door for application-specific software optimizations which can take advantage of specific hardware features of GPUs. 
\vspace{-2mm}
\subsection{Application-Driven Data Prefetching}
\label{subsec:design_prefetching}
Data prefetching is a classic technique to improve performance for memory latency-bound kernels and has been well-adopted in both software~\cite{cpu_prefetch_intrinsics} and hardware~\cite{lee2012prefetching, falsafi2022primer}. However, on-chip prefetching in GPUs is {\em uncommon}, due to GPUs extensive reliance on WLP for latency hiding~\cite{why_gpu}. Also, unlike CPUs, GPUs do not employ dedicated on-chip hardware-based prefetching engines. Thus, the CUDA programmer can develop tailor-made prefetching solutions for their application while minimizing its overhead challenges. To comprehensively explore the design space of prefetching, we formulate a series of key questions as discussed next: 
%(i) What to prefetch?, (ii) Where to prefetch?, (iii) How and When to prefetch?, and (iv) How to counter possible overheads?

\textbf{What to Prefetch?} Recall that Section~\ref{subsec:work_partitioning} highlighted the presence of the gather-reduce operations (forming a load-use dependency chain) as part of pooling operations performed by each CUDA thread. Here, the gather operation is an indirect load (pointer-based load) and spans a variety of memory access patterns (Table~\ref{tab:unique_accesses_per_dataset}), which leads to long scoreboard stalls (Tables~\ref{tab:base_profiling}  and~\ref{tab:OptMT_profiling}). Thus, this gather operation is our prefetch target. During the kernel launch of the embedding bag operator, each CUDA thread receives an offset and indices array. These two arrays essentially provide each thread with the complete set of addresses it will need to load data from. Consequently, by leveraging this knowledge of future access patterns before the actual loads are required, we can insert 100\% accurate prefetches.

\begin{figure}
\centering
\includegraphics[width=0.8\columnwidth]{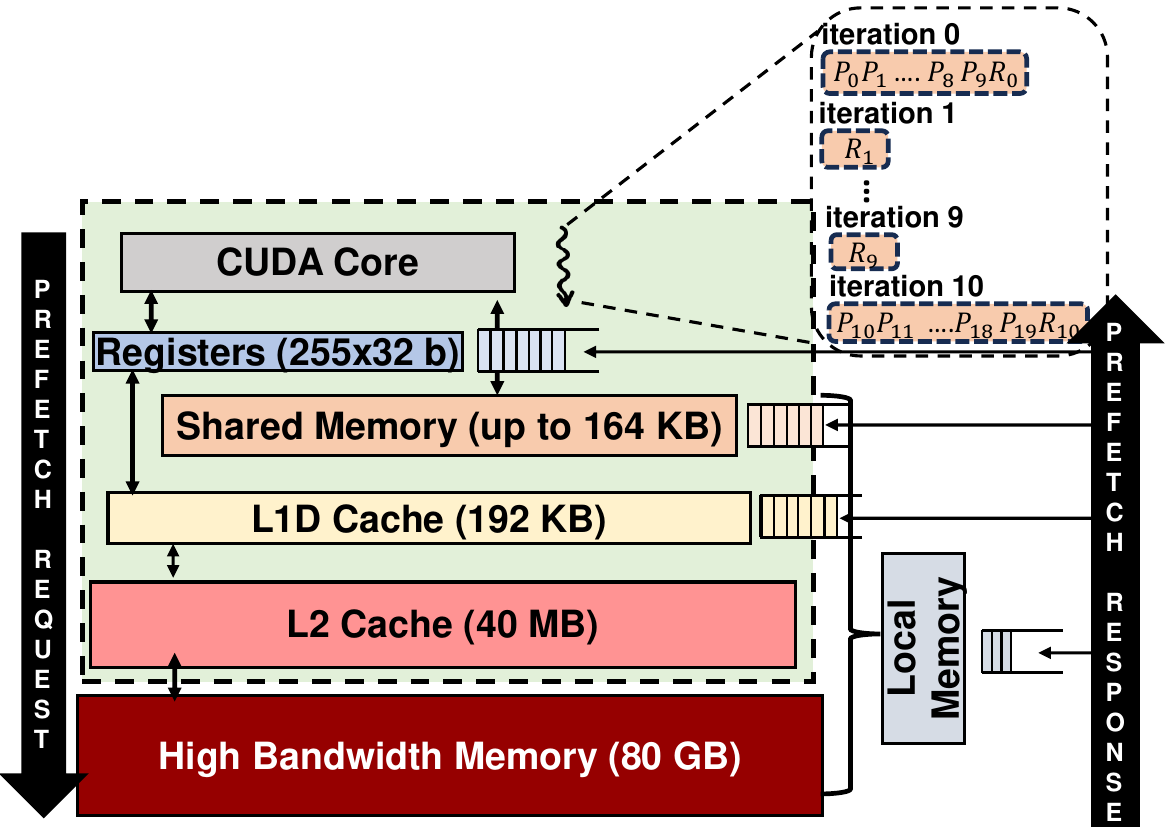}
\caption{Buffer locations used for various prefetching schemes. A CUDA thread is shown executing SMPF, a batch of 10 prefetches(P) are launched every 10th iteration, and a reduce(R) operation every iteration.}
\label{fig:prefetch_levels}
\vspace{-0.2in} 
\end{figure}

\textbf{Where to Prefetch?} Ideally, for the goal of latency hiding, we would like to prefetch the data as close to the CUDA core pipeline as possible. However, the hardware could be limited in resources, and thus it is wise to consider a variety of locations for storing the prefetches (buffer stations). Figure~\ref{fig:prefetch_levels} shows a total of 4 buffer stations -- register, shared memory, L1 D\$, and local memory. We do not pick L2 since its access latency is quite high (261.5 cycles~\cite{luo2024benchmarking}). Note that each buffer station has pros and cons. Specifically, in terms of access latency, the register is optimal, whereas, in terms of size, L1 D\$ and local memory are optimal locations. Note also that local memory is the scope of a variable, and the data can reside in L1/L2/HBM~\cite{local_memory}. We design and implement prefetching for all 4 buffer locations, and use the following abbreviations: RPF for Register-based Prefetching, SMPF for Shared memory-based Prefetching, LMPF for Local memory-
based Prefetching, and L1DPF for L1D\$ Prefetching.
\begin{figure}
\centering
\includegraphics[width=1\columnwidth]{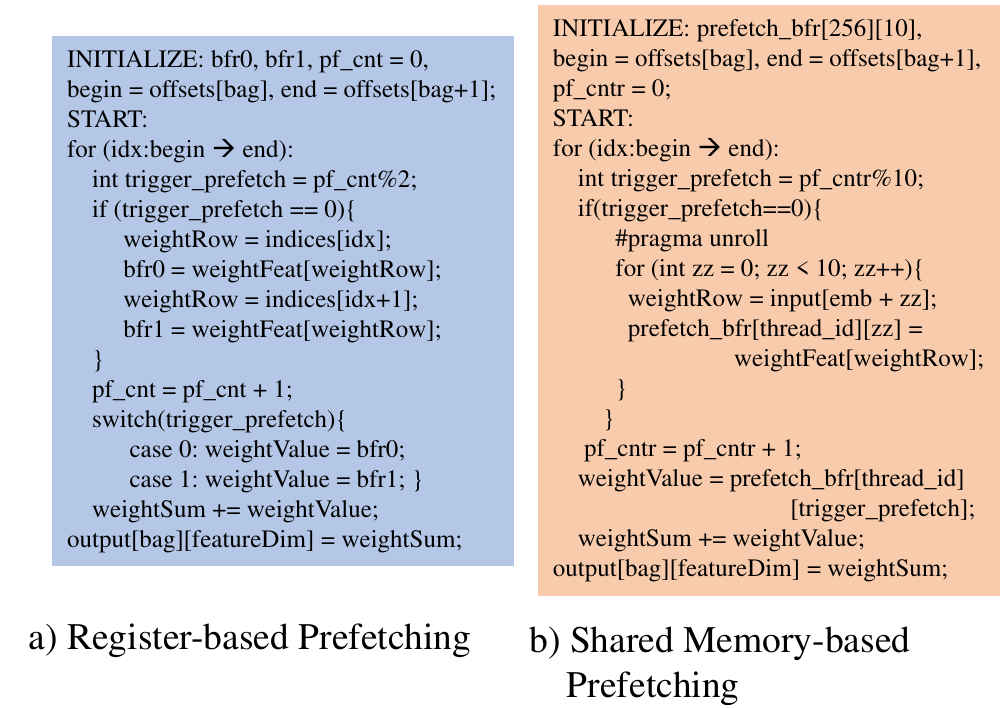}
\caption{Prefetching implementations: (a) RPF (b) SMPF}
\label{fig:prefetch_implementations}
\vspace{-0.2in} 
\end{figure}

\begin{figure}
\centering
\includegraphics[width=0.8\columnwidth]{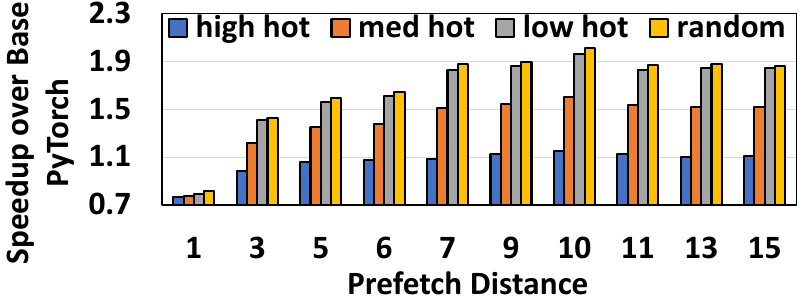}
\vspace{-2mm}
\caption{Performance impact of prefetch distance in SMPF.}
\label{fig:vary_prefetch_distance}
 
\end{figure}

\textbf{How and When to Prefetch?} Fundamentally, we want to issue a prefetch much ahead of its demand to hide the worst-case latency (timeliness property). Also, as mentioned at the beginning of the section, GPU employs a scoreboard mechanism for ILP. In the CUDA program, one can take advantage of it by manually reordering the instruction stream to pack a batch of needed load instructions. Towards this, for RPF, SMPF, \& LMPF, (1) we mimic prefetching with the ahead-of-time issue of the demand loads and storing the data into a buffer station, thus the prefetch becomes the producer of the data, and (2) when this data is consumed by the reduce operation, it fetches it from same buffer station. Based on this understanding, Figure~\ref{fig:prefetch_implementations} shows a simplified implementation of the RPF and SMPF schemes on top of Algorithm 2. LMPF can be similarly implemented. For L1DPF, we use a PTX-based intrinsic ``prefetch.global.L1"~\cite{ptx_prefetch} to prefetch the data into L1D\$, similar to commonly used CPU intrinsics~\cite{cpu_prefetch_intrinsics}. To achieve timeliness (and ultimately optimal performance), we vary the prefetch distance to find the optimal value. Figure~\ref{fig:vary_prefetch_distance} evaluates the prefetch distance for the SMPF scheme over various datasets and finds the optimal prefetch distance as 10. 

\textbf{Countering the Potential Overheads?} 
As discussed above, by systematically navigating the search space, prefetching can find the optimal design points and improve performance. However, various overheads could arise: (1) With the addition of software prefetch support, the total instructions executed could increase, leading to increased computation. For example, a 37.2\% overhead is observed in SMPF. 
%(Figure~\ref{fig:prefetching_comparison_base})
(2) Many prefetch distance choices could degrade the performance. For example, in Figure~\ref{fig:vary_prefetch_distance}, a distance of 1 hurts the performance for all datasets. Also, similar to ~\cite{sethia2015mascar}, a large distance could create LSU stalls. Furthermore, injected prefetches could hurt the locality of other parts of the code.  (3) With the optimal prefetch distance of 10 in Figure~\ref{fig:vary_prefetch_distance}, for random case, while the long scoreboard stalls significanlty reduce from 18.6 cycles(Table~\ref{tab:base_profiling}) to 4.6 cycles, the remaining stalls suggest suboptimal timeliness.  (4) For a given buffer station, we are limited by either the size or latency it offers. For example, while the register access is fastest (Table~\ref{tab:access-latency-gpu-mem}), their limited count can become a bottleneck, as noted with limited WLP and register spilling (Section~\ref{subsec:characterization}).(5) Though not in our case, any scope of inaccurate prefetching or unavailability in off-chip bandwidth headroom could hurt performance by causing cache pollution~\cite{wu2011pacman} or throttling of demand memory requests~\cite{emma2005exploring}.

Therefore, through a combination of profiling-based study and empirical tuning, we establish a high-performance prefetching strategy that delivers 100\% coverage and accuracy. Yet, various overheads (like extra instructions and increase in bandwidth demand) and sub-optimal timeliness could hold. To resolve these, in the next subsection, we propose L2 pinning which can complement prefetching. 

% extra instructions,
% if not an ideal distance, low vs high -- backpressure could happen,
% interference across warps,
% mercy of memory bandwidth,
% Conclude with: for a given machine, profiling based tuning is suggested to find the i deal buffer station, prefetch distance. For any other memory bound kernel, first establish that it is memory latency bound, and locate the problematic load. 

% Applying software prefetching by tailoring towards the application needs and opportunities from the GPU

% Mimicking prefetching by inserting explicit loads with various buffering locations. (give a diagram to illustrate the possible locations: register file, shared memory, L1 data cache, L2 cache, HBM). 

% Give the algorithm for inserting prefetch for both register and shared memory. 

% Answer the questions: what to prefetch, when (give inspiration from ~\cite{jain2023optimizing}, where, how

% Ensure that you have answered the basic questions on time liness, cache pollution, downsides of prefetching.

% Put common things first: what to prefetch
% Ways: register, shared memory, local memory, and L1D\$
% For each way: when to prefetch, where to prefetch, how to prefetch
% Emphasize tuning of prefetching is a design decision

\subsection{Application-Aware L2 pinning}
\label{subsec:design_pinning}

Traditionally, GPUs have prioritized compute cores over cache capacities while relying on WLP for {\em tolerating} long latency stalls. However, with the rise of new applications (e.g., those based on deep learning) and larger chip areas, modern GPUs are witnessing an increase in cache capacities (Table~\ref{tab:gpu-cache-capacities}). Further, Nvidia GPUs (from Ampere architecture onwards) have recently released support for CUDA programmer-based L2 cache access management (residency control~\cite{l2_residency}, which enables a portion of the L2 cache to be used for persistent data access). Given that off-chip memory accesses are very costly (Table~\ref{tab:access-latency-gpu-mem}), a programmer with the knowledge of the underlying memory access pattern behavior can {\em mark} the persistence of high-reuse regions that otherwise may suffer from thrashing. Given that the memory accesses in the embedding stage follow a Power Law  distribution (Figure~\ref{fig:coverage_study}), we propose an L2 pinning (L2P) design that can benefit from L2 residency control for the embedding accesses. Towards this, we first discuss the design and implementation, and then discuss the performance expectations and associated overheads.

\begin{figure}
\centering
\includegraphics[width=0.7\columnwidth]{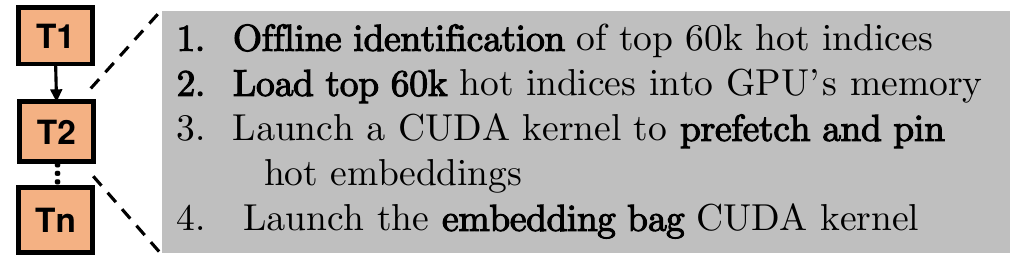}
\caption{High-level design of L2P}
\label{fig:L2_pinning_steps} 
\end{figure}

\textbf{What, When, and How to Pin?} In an A100 GPU, a maximum of 30MB (75\% of L2) can be set-aside for residency control, while the remaining (at least 10MB) is completely hardware managed. To promote the highest locality, we propose using the complete 30MB for storing the most frequent embedding vectors. Since each embedding vector is of size 512KB, thus a maximum of 60K embedding vectors can be pinned in L2. Figure~\ref{fig:L2_pinning_steps} shows the high-level design for L2P. We conduct an offline profiling to identify the top 60K hot indices present for each dataset which are used as candidate embedding entries for pinning. In the beginning of the inference server, these indices can be loaded into the GPU's main memory. The embedding tables are processed sequentially, and each table follows two steps: (1) launch a CUDA kernel that prefetches and pins embedding vectors corresponding to the hot indices, and (2) launch the default embedding bag CUDA kernel. We use the inline PTX instruction "prefetch.global.L2::evict\_last"~\cite{ptx_prefetch},  which takes an address as input, loads the associated cache line into L2, and marks the eviction policy as ``evict\_last". Setting this eviction policy allows the marked/chosen data to persist in L2 over others. For instance, during an eviction event in L2 for a set, the marked cache lines are less likely to be kicked out. Thus, the embedding stage is expected to observe lower access latency for a majority of the memory accesses (261 cycles over 466 cycles from Table~\ref{tab:access-latency-gpu-mem}).

\begin{figure}
\centering
\includegraphics[width=0.7\columnwidth]{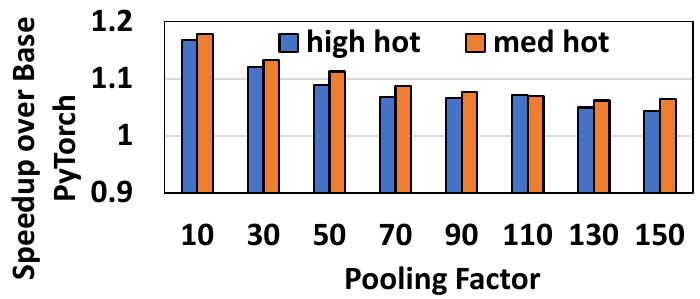}
\caption{Detailed study of L2 pinning over various pooling factors. Speedup is reported over off-the-shelf PyTorch.}
\label{fig:l2_pinning_base}
\vspace{-0.2in} 
\end{figure}

\textbf{Performance Expectations and Overheads?} 
Although hardware caches capture hotness to an extent (Table~\ref{tab:base_profiling}), we expect L2 pinning to further enhance locality by avoiding: (1) L2 cold misses for hot indices, and (2) thrashing of highly reused embeddings. Moreover, the effectiveness of hardware caches is impacted by the batch and pooling sizes (based on the model and scheduling policy) where smaller sizes lead to lower reuse situations. As expected, L2P improves over the baseline, yielding more benefits in lower pooling cases, and performs slightly better for the med hot case (Figure~\ref{fig:l2_pinning_base}).  

Clearly, embedding access patterns can change over time, potentially reducing the effectiveness of L2 pinning. To address this challenge, similar to prior research~\cite{lee2021merci}, we can update the pinned data periodically. This ensures that the L2 cache always stores the most frequently accessed elements, maximizing the benefit of pinning. The overhead of storing the top 60K indices for every table on a GPU is minimal. For example, for 250 tables, it would be 250$\times$60K$\times$8B = $\sim$120MB. Finally, the overhead of the L2P kernel is small and can be hidden by overlapping it with the CPU pre-processing required before the embedding bag kernel launch.

% Shortcomings of prefetching on top of which pinning is needed.
% Pinning opportunity arises with Ampere architecture.
% Pinning design
% Counter overheads
\vspace{-2mm}
\subsection{Synergy between Prefetching and Pinning}
\label{subsec:design_synergy}
In this subsection, we discuss how prefetching and pinning can {\em symbiotically} work with each other, and thus further improve upon embedding bag's memory-latency bound regime. While prefetching hides the long load latency by bringing the demand loads ahead of time near the CUDA core pipeline (registers, shared memory, or L1D\$), it still suffers from suboptimal timeliness and puts pressure on the off-chip bandwidth. Similarly, even though pinning lowers the load latency by bringing and holding the frequently-accessed embeddings in the L2 cache, it still lacks because  (1) 30MB of L2 set-aside cannot fully cover the working set required by the datasets, especially in the low hot and random cases, and (2) access latency with L2 is significantly high (261.5 cycles; see Table~\ref{tab:access-latency-gpu-mem}), compared to registers, shared memory, or L1D\$. When combined, prefetching strengthens pinning by providing 100\% coverage and faster access to the CUDA core pipeline, and pinning bolsters prefetching by improving the timeliness while cutting down on HBM requests.

%%%%%

% Benefit for Pinning: data is getting closer to demand load
% Benefit for prefetching: Timeliness of prefetching as latency of L2 is better than global, reduction in L2 bandwidth usage.

% TODO:
% \begin{itemize}
%   \item DONE Previous section established memory latency as a critical problem. What can we do? 
%   \item Could we tailor commonly known software solutions for embedding bag CUDA kernel? 
%   \item DONE do highlight the properties of baseline: compiled using nvcc with O3 flag, naive loop unrolling doesn't help in reordering instructions. 
%   \item DONE discuss prefetching, pinning. Highlight that prefetching is not commonly used in GPUs. Also discuss the status of preference in deployment of memory bound workloads on CPUs vs GPUs (like the ones with low arithmetic intensity).
%   \item TODO while discussing designs, highlight the novelty again
%   \item DONE Application-specific Software Prefetching
%   \item DONE Application-specific L2 Pinning
%   \item DONE each design should mention: (1) tradeoffs on benefit vs cost (2) analytically, what could one expect in the evaluation before even looking at the empirical experiment results.
  
% \end{itemize}
\section{Methodology}
\label{sec:methodology}
\textbf{Hardware:} %Nvidia's A100 GPU is used in our study. 
Table \ref{tab:hardware} captures the hardware properties of our real-system evaluation setup. 

\begin{table}[!ht]
\caption{System specifications used for evaluation }
\vspace{-2mm}
    \centering
    %\footnotesize
    \scriptsize
    \setlength{\tabcolsep}{6pt}
    \begin{tabular}{|c|c|}
    \hline
        \textbf{CPU} & AMD EPYC 7763\\ \hline
        \textbf{RAM} & 1 TB\\ \hline
        \textbf{GPU} & Nvidia A100-SXM4-80GB\\ \hline
        \textbf{\# SMs} & 108\\ \hline
        \textbf{Register File per SM} & 64K $\times$ 32 bit\\ \hline
        \textbf{L1D Cache size} & 192KB\\ \hline
        \textbf{Shared Memory size} & up to 164KB\\ \hline
        \textbf{L2 Cache size} & 40MB\\ \hline
        \textbf{Device Memory} & 80GB, HBM2e\\ \hline
        \textbf{HBM Bandwidth} & 1.94TB/s\\ \hline
       
    \end{tabular}
    
    \label{tab:hardware}
\vspace{-0mm}
\end{table}

\textbf{Software:} PyTorch (v2.1.0) ~\cite{pytorch_2.1} is source-compiled on a Linux/Ubuntu machine with CUDA Driver Version: 535.129.03, and nvcc version 12.2. Here, source-compiled PyTorch matches off-the-shelf packaged PyTorch in performance. ``nvcc"  compilation is highly optimized with O3 flag~\cite{nvcc_compiler}. %All schemes have been checked for functional correctness.

% \textbf{Profiling:} \todo{TODO Rishabh}

\textbf{Model:} Taking inspiration from ~\cite{gupta2020architectural, gupta2020deeprecsys, mudigere2022software, jain2023optimizing}, we pick model configurations which are representative of industrial inference settings.  
%Our considered DLRMs are similar to embedding-heavy models mentioned in ~\cite{gupta2020deeprecsys, jain2023optimizing, hsia2020cross} 
Following are used: (1) Bottom MLP dimensions are "1024-512-128-128" (2) Embedding stage has 250 tables, each table having 500,000 rows and 128 embedding dimensions (3) Top MLP dimensions are "128-64-1". 4-byte precision is used which makes each embedding vector of size 512KB. The total model weight is of size $\sim$60 GB which can completely fit within one GPU's memory while the remaining memory is used for computations across intermediate layers. Unless mentioned otherwise, all tables are homogeneous in hotness. Table~\ref{tab:heterogeneous} describes a mixture of tables for heterogeneous evaluation.

\begin{table}[!ht]
\caption{Heterogeneous Mixture of Model configuration}
\vspace{-2mm}
    \centering
    %\footnotesize
    \scriptsize
    \setlength{\tabcolsep}{6pt}
    \begin{tabular}{|c|c|c|c|c|}
    \hline
        \textbf{Mixture/Datasets} &\textbf{high hot}&\textbf{med hot}&\textbf{low hot}&\textbf{random}\\ \hline
       \textbf{Mix1 (\#tables)} & 100 & 75 & 50 & 25 \\ \hline
       \textbf{Mix2 (\#tables)} & 62 & 63 & 63 & 62 \\ \hline
       \textbf{Mix3 (\#tables)} & 25 & 50 & 75 & 100 \\ \hline
    \end{tabular}
    
    \label{tab:heterogeneous}
\vspace{-0mm}
\end{table}

\textbf{Datasets:} Following an earlier work ~\cite{jain2023optimizing}, we use the publicly released homogenized production traces from Meta ~\cite{homogeneous-dlrm-dataset, dlrm-dataset}. Thus, we consider a variety of hotness (memory access patterns): one item, high hot, medium hot, low hot, and random. Section ~\ref{subsec:memory_access_patterns} quantitatively compares these datasets.  To represent a large access pool similar to a real inference, we calculate the unique access \% averaged over 100 measurements (Table~\ref{tab:unique_accesses_per_dataset}). Inspired by ~\cite{sethi2022recshard, gupta2020deeprecsys, jain2023optimizing, dlrm-dataset}, a large batch size of 2048 and a large ``lookups per sample" (or pooling factor) of 150 is used, . 

\textbf{Nomenclature for combined schemes:}
% As described in Section~\ref{subsec:characterization}, OptMT refers to Optimal Multi-threading scheme which provides (ideally) 40 active warps per SM. In the prefetching schemes (Section~\ref{sec:optimizations}), RPF refers to Register-based Prefetching, SMPF refers to Shared memory-based Prefetching, LMPF refers to Local memory-based Prefetching, and L1DPF refers to L1D\$ intrinsic-based Prefetching. L2P refers to L2 Pinning scheme. 
Any combined scheme is denoted using a `+' symbol. For example, RPF+L2P+OptMT is a combination of 3 schemes, namely, RPF, L2P, and OptMT. 
\vspace{-2mm}

% \begin{itemize}
%   \item Hardware details: in a table (if 3 GPUs, it could be a nice table)
%   \item Software details: Latest PyTorch 2.1.0, latest CUDA
%   \item give details on running the complete model on GPU: stating that all the tables execute on the GPU where intermediate data stays on the GPU itself.
%   \item DLRM model and dataset (give link to my ISCA paper for the datasets)
% \end{itemize}
%%%%%

\section{Evaluation}
\label{sec:evaluation}
Based on real system measurements, we evaluate the benefits of our proposed latency hiding schemes: OptMT (Section~\ref{subsec:characterization}), Software Prefetching [RPF, SMPF, LMPF, L1DPF], L2 Pinning (L2P), as well as their combined versions (Section~\ref{sec:optimizations}).
%(recall that the associated methodology was described earlier in Section~\ref{sec:methodology}).
We first study the key improvements in embedding stage and end-to-end DLRM with micro-architectural justifications, followed by sensitivity studies. In all our evaluations below, off-the-shelf PyTorch~\cite{emb_bag_cuda_kernel, pytorch_dlrm} (with the property of 24 theoretical active warps per SM) is used as the ``baseline", and all the stages of DLRM inference run on a GPU. Performance is measured as ``batch latency", and all improvements are reported over off-the-shelf/base PyTorch.
\vspace{-2mm}
\subsection{Key Results}
\subsubsection{Boost in Embedding Stage Performance}
\begin{figure}
\centering
\includegraphics[width=0.8\columnwidth]{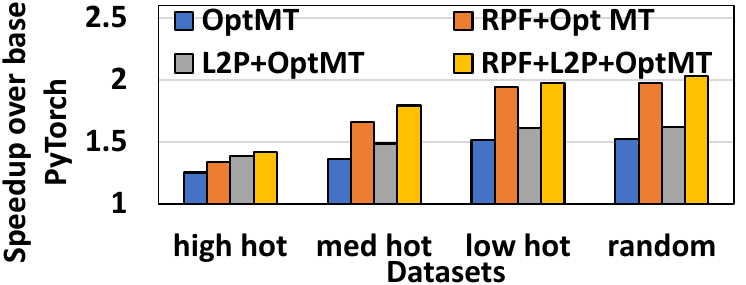}
\caption{Embedding-only improvement in latency of the proposed techniques with OptMT over off-the-shelf PyTorch.}
\label{fig:emb_only}
\vspace{-0.0in} 
\end{figure}

Given our proposed schemes target the primary bottleneck (Embedding Stage) in DLRM, we first highlight the embedding-only benefits. Figure~\ref{fig:emb_only} evaluates the performance of various design points across different datasets. As highlighted earlier in Section~\ref{subsec:characterization}, OptMT exploits the GPU's WLP better, thus, better hiding the long latency loads. RPF and L2P complement OptMT by further lowering the latency (note that we have picked RPF as the winning prefetching scheme due to it performing slightly better over other schemes; see  Section~\ref{subsubsec:winning_prefetching}). As L2P is more useful for smaller working set situations, more improvement is seen with the high/med hot cases, and as RPF is more suitable for long latency load situations, more improvement is seen with low hot/random cases. Finally, we note that RPF and L2P combined further improves the performance, achieving up to a 2.03$\times$ speedup (random case). Interestingly, the highest benefit is 13.5\% in the med hot case over the previous optimal (RPF+OptMT), thus maximally complementing each other.

\subsubsection{Boost in End-to-End DLRM Performance}
\begin{figure}
\centering
\includegraphics[width=0.8\columnwidth]{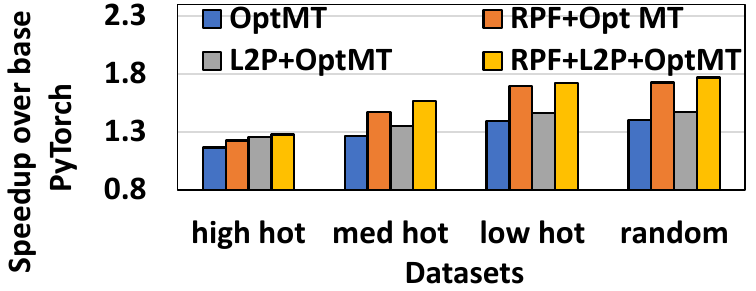}
\caption{End-to-end improvement in latency of the proposed techniques with OptMT over off-the-shelf PyTorch.}
\label{fig:endtoend}
\vspace{-0.2in} 
\end{figure}

\begin{figure}
\centering
\includegraphics[width=0.6\columnwidth]{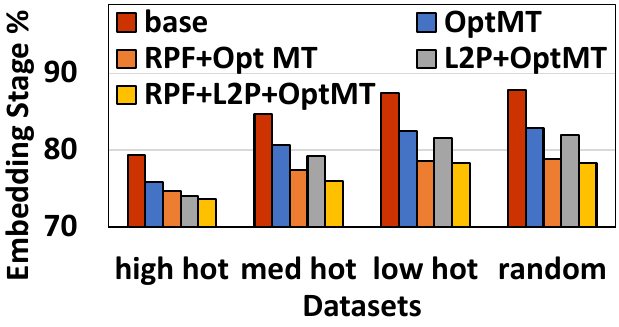}
\caption{Embedding stage contribution in the end-to-end latency.}
\label{fig:emb_contribution}
\vspace{-0.2in} 
\end{figure}

Given that DLRM is composed of 4 stages (Section~\ref{subsec:model_arch}) which collectively influence the batch latency, we also evaluate the benefit of our proposed schemes with respect to the end-to-end latency (Figure~\ref{fig:endtoend}). Note that the trends in speedup remain similar to Figure~\ref{fig:emb_only}, with a minor degradation in the final speedups (since embedding is the bottleneck). It can also be noted that the combined scheme (RPF+L2P+OptMT) achieves a significant speedup of up to 77\% (random case). 
%Figure~\ref{fig:emb_contribution} highlights the embedding stage contribution in the end-to-end latency for all datasets and schemes \textcolor{red}{Highlight some trend in the figure since you are bringing it up here}. 
Recall Figure~\ref{fig:intro} highlighted the performance gap between the fastest (one item dataset) and slowest (random dataset) loads being 3.2$\times$ and 2.1$\times$ for off-the-shelf and OptMT Pytorch, respectively. With the synergistic integration of our proposed schemes in play, we are able to substantially lower the performance gap (higlighted in Figure~\ref{fig:emb_contribution}) to only 1.57$\times$, thus decreasing it by 163\% and 53\% over off-the-shelf and OptMT PyTorch, respectively. Therefore, with the Embedding Stage running optimally, its contribution in the end-to-end execution reduces by up to 10\% (random). 

\subsubsection{Microarchitectural Justifications}
To better understand the above gains, we profile the proposed schemes using NCU~\cite{ncu_tool}. \textcolor{black}{Table~\ref{tab:RPF+OptMT_profiling} and Table~\ref{tab:RPF+L2P+OptMT_profiling} show the microarchitectural measurements for RPF+OptMT and RPF+L2P+OptMT designs, respectively. Due to limitations in NCU  profiling~\cite{ncu_profiling_limitation}, we could not measure all the metrics for the integrated scheme.} For the random case, RPF+L2P+OptMT achieves an issue slot utilization of \textcolor{black}{44\%}, thereby improving by \textcolor{black}{83\%} over baseline, and \textcolor{black}{26\%} over OptMT. This is because RPF+OptMT better utilizes the memory bandwidth, reaching up to \textcolor{black}{700} GBps, significantly higher than the baseline (329.5 GBps). With L2P combined, for the high and med hot cases, it lowers the total amount of data read from the device memory by \textcolor{black}{71}\% and \textcolor{black}{16.2}\%, thus lowering memory access latencies \textcolor{black}{and saving memory bandwidth.} 

\begin{table}[!ht]
\caption{\textcolor{black}{Microarchitectural details for RPF+OptMT}}
    \centering
    %\footnotesize
    \scriptsize
    \setlength{\tabcolsep}{6pt}
    
    %\begin{tabular}{|c|c|c|c|c|c|}
    \begin{tabular}{|p{3.8cm}|p{0.4cm}|p{0.4cm}|p{0.4cm}|p{0.4cm}|p{0.6cm}|}
    \hline

    \textcolor{black}{\textbf{NCU metrics/datasets}}                            & \textcolor{black}{\textbf{high hot}} & \textcolor{black}{\textbf{med hot}} & \textcolor{black}{\textbf{low hot}} & \textcolor{black}{\textbf{rand}} \\ \hline
    \textcolor{black}{\textbf{Kernel time (us)}}                    & \textcolor{black}{177} & \textcolor{black}{205} & \textcolor{black}{220} & \textcolor{black}{224} \\ \hline 
    \textcolor{black}{\textbf{\#load insts (M)}}                    & \textcolor{black}{4.43}     & \textcolor{black}{4.43}     & \textcolor{black}{4.43}     & \textcolor{black}{4.43}     \\ \hline
    \textcolor{black}{\textbf{SM Throughput \%}}                & \textcolor{black}{59.3}    & \textcolor{black}{49.7} & \textcolor{black}{44.4}    & \textcolor{black}{43.3}     \\ \hline
    \textcolor{black}{\textbf{issued slot utilization (\%)}}            & \textcolor{black}{59.17}     & \textcolor{black}{49.65}     & \textcolor{black}{44.32}     & \textcolor{black}{43.5}     \\ \hline
    \textcolor{black}{\textbf{Device Memory size read(MB)}}                   & \textcolor{black}{8.4}      & \textcolor{black}{53}     & \textcolor{black}{133}    & \textcolor{black}{151.8}    \\ \hline
    \textcolor{black}{\textbf{Avg HBM Read BW(GBps)}}              & \textcolor{black}{51.4}       & \textcolor{black}{277.7}    & \textcolor{black}{629.1}    & \textcolor{black}{699.4}   \\
    \hline
    \textcolor{black}{\textbf{Avg HBM Read BW Utilization (\%)}} &\textcolor{black}{2.6}                      & \textcolor{black}{13.9}              & \textcolor{black}{31.5}           & \textcolor{black}{35} \\\hline

    \end{tabular}
    
    \label{tab:RPF+OptMT_profiling}
\vspace{-2mm}
\end{table}  

\begin{table}[!ht]
\caption{\textcolor{black}{Microarchitectural details for RPF+L2P+OptMT}}
    \centering
    %\footnotesize
    \scriptsize
    \setlength{\tabcolsep}{6pt}
    
    %\begin{tabular}{|c|c|c|c|c|c|}
    \begin{tabular}{|p{3.8cm}|p{0.4cm}|p{0.4cm}|p{0.4cm}|p{0.4cm}|p{0.6cm}|}
    \hline

    \textcolor{black}{\textbf{NCU metrics/datasets}}                            & \textcolor{black}{\textbf{high hot}} & \textcolor{black}{\textbf{med hot}} & \textcolor{black}{\textbf{low hot}} & \textcolor{black}{\textbf{rand}} \\ \hline
    \textcolor{black}{\textbf{Kernel time (us)}}                    & \textcolor{black}{167} & \textcolor{black}{190} & \textcolor{black}{216} & \textcolor{black}{217} \\ \hline 
    \textcolor{black}{\textbf{\#load insts (M)}}                    & \textcolor{black}{4.43}     & \textcolor{black}{4.43}     & \textcolor{black}{4.43}     & \textcolor{black}{4.43}     \\ \hline
    \textcolor{black}{\textbf{SM Throughput \%}}                & \textcolor{black}{60}    & \textcolor{black}{49.9} & \textcolor{black}{44.5}    & \textcolor{black}{43.3}     \\ \hline
    \textcolor{black}{\textbf{issued slot utilization (\%)}}            & \textcolor{black}{60.12}     & \textcolor{black}{50.21}     & \textcolor{black}{44.64}     & \textcolor{black}{43.61}     \\ \hline
    \textcolor{black}{\textbf{Device Memory size read(MB)}}                   & \textcolor{black}{4.9}      & \textcolor{black}{45.6}     & \textcolor{black}{128}    & \textcolor{black}{150}    \\ \hline
    \textcolor{black}{\textbf{Avg HBM Read BW(GBps)}}              & \textcolor{black}{30}       & \textcolor{black}{240.6}    & \textcolor{black}{613.2}    & \textcolor{black}{698}   \\
    \hline
    \textcolor{black}{\textbf{Avg HBM Read BW Utilization (\%)}} &\textcolor{black}{1.5}                      & \textcolor{black}{12.3}              & \textcolor{black}{30.7}           & \textcolor{black}{34.9} \\\hline

    \end{tabular}
    
    \label{tab:RPF+L2P+OptMT_profiling}
\vspace{-2mm}
\end{table}

% Highlight: (1) benefits from register file prefetching in bandwidth, and compute utilization. (2) benefits of L2 pinning in reducing the total bytes read from device memory. 
\vspace{-2mm}
\subsection{Sensitivity Analysis}
\subsubsection{Winning Prefetching Scheme}
\label{subsubsec:winning_prefetching}
\begin{figure}
\centering
\includegraphics[width=0.7\columnwidth]{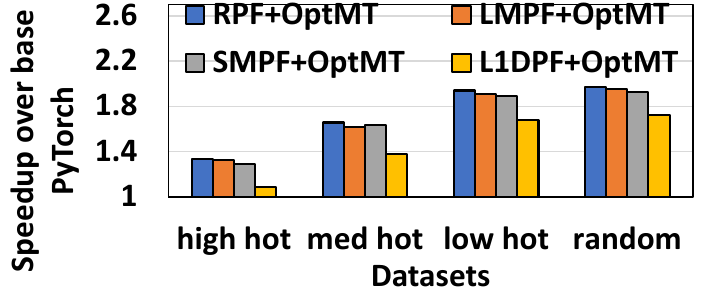}
\caption{Comparison of all prefetching techniques with OptMT over off-the-shelf PyTorch.}
\label{fig:prefetching_comparison}
\vspace{-0.2in} 
\end{figure}

Earlier, Section~\ref{sec:optimizations} proposed 4 different data prefetching schemes based on the buffering location of the prefetches -- register, shared memory, local memory, or L1 data cache. Because of this, the implementation of each scheme differs. For each prefetching scheme (on top of OptMT), we empirically find the optimal ``prefetch distance" by doing a sweep similar to that in Figure~\ref{fig:vary_prefetch_distance}. Interestingly, we find that all schemes perform best at a prefetch distance of 2. Figure~\ref{fig:prefetching_comparison} compares all prefetching schemes in conjunction with OptMT over the baseline PyTorch. It can be noted that all schemes improve on top of the baseline, where L1DPF improves the least and RPF improves the most. However, to implement prefetching, we modify the CUDA kernel which results in extra instructions, thus adding overhead by increasing the total raw instructions to be processed. Among all the schemes tested, we note that L1DPF suffers the most from this overhead. We further note this overhead becomes more critical in the high hot cases where the prefetching finds less opportunity to benefit, resulting in a 15\% drop in speedup when compared to the only OptMT.

RPF, SMPF, and LMPF perform very close in all the datasets,  with RPF marginally winning. This is because the register file is closest to the execution pipeline when compared to other buffer locations (Table~\ref{tab:access-latency-gpu-mem}). Thus, prefetching achieves \{34\%, 66\%, 94\%, 97\%\} speedups for the \{high, med, low\} hot and random, respectively. 

% \begin{figure}
% \centering
% \includegraphics[width=0.8\columnwidth]{figures/l2_pinning_OptMT.pdf}
% \caption{Detailed study of L2 pinning with OptMT over various pooling factors. Speedup is reported over off-the-shelf PyTorch.}
% \label{fig:l2_pinning_OptMT}
% \vspace{-0.2in} 
% \end{figure}

\subsubsection{Improvement over Baseline PyTorch without OptMT}
\iffalse
\begin{figure}
\centering
\includegraphics[width=0.8\columnwidth]{figures/prefetching_comparison_base.pdf}
\caption{Comparison of all prefetching techniques over off-the-shelf PyTorch.}
\label{fig:prefetching_comparison_base}
\vspace{-0.2in} 
\end{figure}

\begin{figure}
\centering
\includegraphics[width=0.8\columnwidth]{figures/emb_only_base.pdf}
\caption{Embedding-only improvement of the proposed techniques over off-the-shelf PyTorch.}
\label{fig:emb_only_base}
\vspace{-0.2in} 
\end{figure}
\fi
\begin{figure}
\centering
\includegraphics[width=0.8\columnwidth]{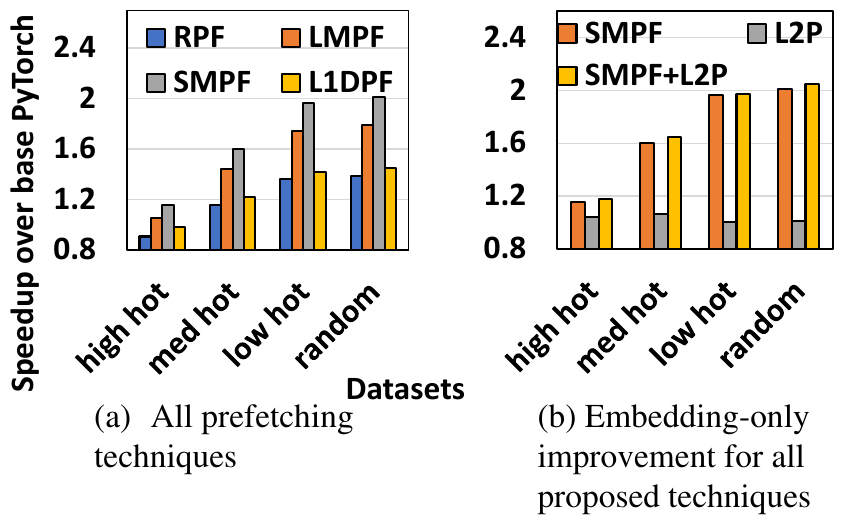}
\caption{Comparison of techniques for off-the-shelf Pytorch.}
\label{fig:prefetching_comparison_base}
\vspace{-6mm} 
\end{figure}

Earlier, Figure~\ref{fig:vary_wlp} highlighted that OptMT improves the performance over baseline PyTorch, and our proposed techniques can work in conjunction (Figure~\ref{fig:emb_only}) with OptMT. However, we also evaluate our proposed schemes directly over baseline (no OptMT) to validate their effectiveness in the original situation. Here, we compare all prefetching schemes with their optimal prefetch distance (i.e., \{4,10,10,5\} for \{RPF, SMPF, LMPF, and L1DPF\}, respectively) (Figure~\ref{fig:prefetching_comparison_base}.a). Compared to Figure~\ref{fig:prefetching_comparison}, the winning scheme is not RPF but SMPF. Further, LMPF performs second, L1DPF third, and RPF fourth. For the high hot case, RPF and L1DPF underperforms due to higher instruction overheads. Also, SMPF enhances the performance of all the datasets. When comparing SMPF to RPF+OptMT, it matches the performance in the low hot and random datasets, and slightly underperforms in the high and med hot datasets. We noticed that nvcc compiles the SMPF implementation with 32 warps per SM, instead of 24 (as used in off-the-shelf PyTorch). Using the higher benefit of multi-threading (Figure~\ref{fig:vary_wlp}), this makes SMPF perform better than LMPF, thus making it winning scheme. In contrast, for RPF, nvcc allocates more registers per kernel as the prefetch distance increases, leading to very few (16) warps per SM (for distances $>=$ 5), thus severely hurting performance.

%For RPF, with a prefetch distance of 5 or more, the default nvcc register allocation also increases, thus, building register pressure and leading to (even) 16 warps per SM. 

Considering the winning scheme to be SMPF, Figure~\ref{fig:prefetching_comparison_base}.b, highlights the embedding-only improvement of L2P and SMPF+L2P over the base PyTorch. L2P improves the high hot and med hot cases by 4.5\% and 6.4\%, respectively, while marginally improving for the low hot and random cases. Further, L2P combines with SMPF and further enhances the performance over SMPF only. When comparing to the benefit coming with OptMT (Figure~\ref{fig:emb_only}), it can be seen that SMPF + L2P matches the performance for the low hot and random datasets while slightly underperforming for the high hot and med hot cases. This is because OptMT helps in hiding the latency without adding any instruction overhead. It is also interesting to note that L2P+OptMT performs quite well over OptMT, since L2P better holds the data which otherwise could get evicted due to thrashing across warps.

\begin{figure}
\centering
\includegraphics[width=0.7\columnwidth]{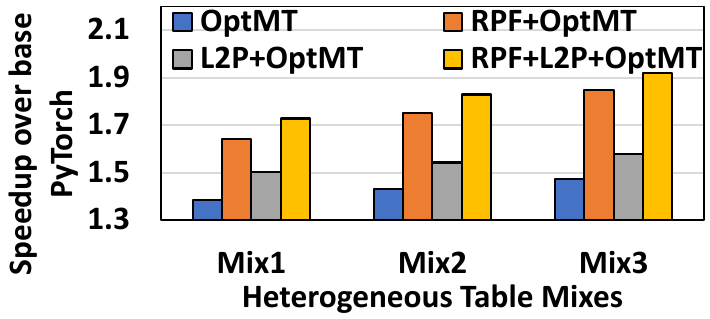}
\caption{Embedding-only improvement of the proposed techniques with OptMT over off-the-shelf PyTorch and heterogeneous tables.}
\label{fig:heterogeneous_mixes}

\end{figure}

\subsubsection{Heterogeneous Table Mixing}
Given that DLRMs are executed in both homogeneous~\cite{gupta2020deeprecsys, jain2023optimizing} and heterogeneous settings~\cite{kal2021space, sethi2022recshard}, we evaluate a synthetic case where non-uniform tables (Table~\ref{tab:heterogeneous}) are present within the embedding stage by having a mixture of hotness. Figure~\ref{fig:heterogeneous_mixes} compares the performance of all proposed schemes in association with OptMT. In general, all schemes perform better on a higher mix due to more contributions from the low hot and random datasets. Further, within any mix, the combined scheme performs the best by improving over any individual scheme.
\vspace{-0mm}

~\textcolor{black}{\subsubsection{Evaluation on H100 NVL GPU}
We also evaluate the applicability our proposed schemes on an H100-NVL GPU ~\cite{h100_nvl}, which is increasingly being embraced by the datacenters ~\cite{meta_h100_infra, amazon_h100}. H100 NVL has 132 SMs (with a total of 16896 CUDA cores), 192KB L1, 50MB L2, 3.84 TBps HBM3 (at 2.7 GHz DDR). The measured base-PyTorch latency values (in us) are \{174, 228, 282, 295\} for the \{high, medium, low, random\} datasets, respectively. Thus, H100 gives an average 47\% uplift in performance (comparing with  (Table~\ref{tab:base_profiling})). Notably, our optimization designs on A100 perform 23\% faster than the H100 base performance, thus making it a more cost-effective solution than simply adopting more expensive GPUs.}    

\begin{figure}
\centering
\includegraphics[width=0.9\columnwidth]{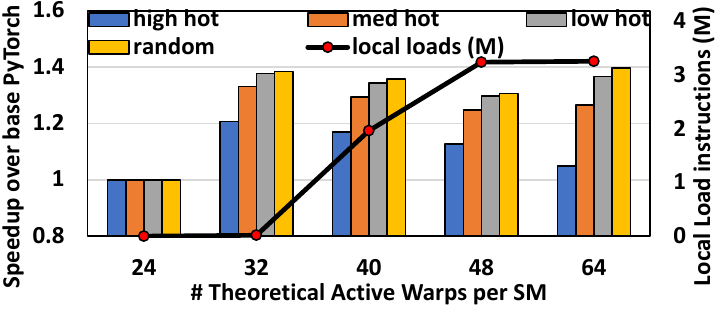}
\caption{\textcolor{black}{On H100 NVL GPU, the number of registers allocated is varied to find optimal WLP. The primary y-axis is speedup over off-the-shelf PyTorch and the secondary y-axis is the register spilling
penalty based on extra local memory loads (in millions).
OptMT on H100 refers to the highest speedup at 32 warps.}}
\label{fig:vary_wlp_h100}
\vspace{-0mm} 
\end{figure}

\begin{figure}
\centering
\includegraphics[width=0.97\columnwidth]{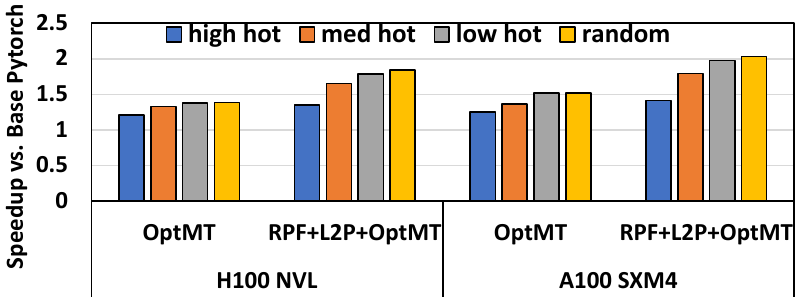}
\caption{\textcolor{black}{Comparison of Embedding-only improvement in latency of the integrated scheme over off-the-shelf PyTorch for H100 NVL and A100 GPU.}}
\label{fig:emb_only_h100}
\vspace{-0mm} 
\end{figure}

~\textcolor{black}{Figure~\ref{fig:vary_wlp_h100} sweeps through possible WLP configurations and finds maximum gain at 32 resident warps (which is different from 40 warps for A100). Similar to Figure~\ref{fig:vary_wlp}, higher gains of MT are visible for low hot and random cases. Finally, Figure~\ref{fig:emb_only_h100} shows the performance benefit of RPF+L2P+OptMT for H100 and compares it with A100. For both OptMT and integrated schemes, H100 observes a little lower speedup compared to A100 which is due to the microarchitectural differences, particularly with H100 having 27\% faster SM clock, 33\% larger L1D\$, 25\% larger L2\$, 20\% wider HBM width, and 64\% faster HBM clock. Yet, we continue to see significant speedups for all datasets (up to 84\%).}

% TODO
% \begin{itemize}
%   \item Only embedding stage improvement: 
%   \item End to End improvement with all 4 stages running on a GPU.
%   \item microarchitecture improvements: compute util, issue slot utilization, bandwidth, cache hit rates, long scoreboard stall, etc.
%   \item evaluating multiple GPUs: A100 vs A6000 vs RTX 3090
%   \item sensitivity study: L2 pinning sweep with PF, BS, ED
%   \item sensitivity study: vary prefetching types register vs shared mem vs local memory vs L1D\$ intrinsic
%   \item sensitivty study: mixed hotness evaluation: eg: 30\% high hot, 30\% MH, 20\%LH, 20\%R​
  
%   % \item Operations Embedding Bag operator: Sum, mean, Max​
%   % \item SLA latencies: latency bounded thpt and tail latency
  
% \end{itemize}

%%%%%

\section{Discussion}
\textcolor{black}{~\textbf{A Static Profiling Framework for adoption of our proposed designs:} Achieving the ideal performance (like Figure~\ref{fig:emb_only}) requires finding the optimal design points for multi-threading and prefetching (where and when to prefetch). We propose a static profiling framework instead of developing an analytical model or heuristics. As the default hardware involves various in-place optimizations like memory-level-parallelism in conjunction with multi-threading, it makes it difficult to holistically capture the complexities with a heuristic. Additional challenges arise with proprietary nvcc and limited public details on GPU microarchitecture, making the heuristics susceptible to errors.} 

%\textcolor{blue}{\emph{(a)Why static profiling is needed?} 
%Achieving the ideal performance (like Figure~\ref{fig:emb_only}) requires finding the ideal design points for the multi-threading, and prefetching (where and when to prefetch). We propose a static profiling framework instead of developing an analytical model. As the default hardware involves various optimizations in-place like memory-level-parallelism in conjunction with multi-threading, it makes it difficult to holistically capture the complexities with a heuristic. Additional challenges arise with proprietary nvcc, and limited public details on GPU microarchitectural.}

%\textcolor{blue}{\emph{(a)Why static profiling is needed?} 

%\textcolor{blue}{\emph{(b) A static profiling framework for finding optimal design points:} 
\textcolor{black}{Thus, we introduce a static profiling framework aimed at conducting design space exploration to identify the most effective design points. The steps are as follows: (i) Assess if the kernel is memory latency bound by checking the memory access patterns, cache misses, and long latency scoreboard stalls. (ii) Assess if the kernel occupancy is maximum. If not, check the usage of register, shared memory, and kernel launch configurations. (iii) If register usage is high, OptMT can be found by varying the allocated registers. Use nvcc compiler flag “-maxrregcount” to control the assigned registers, where the needed registers are $\leq (max\_registers\_per\_SM)  /  ( (desired\_active\_warps) * (warp\_size))$. (iv) After applying OptMT, assess if the kernel is still memory latency bound. If yes, carefully-tuned pinning and prefetching can help. (v) Assess if there is scope for applying L2 pinning by checking for any high reuse behavior towards certain data accesses and comparing the working footprint of kernel with the L2 cache size. If yes, sort the data addresses in descending order based on their reuse amount and apply the steps shown in Figure~\ref{fig:L2_pinning_steps}. (vi) If the performance is still memory latency bound and memory bandwidth is not saturated (under 80\% usage), use prefetching and evaluate performance for different buffer locations as guided in  Figure~\ref{fig:prefetch_implementations} by sweeping across the prefetch distances. Note that, when the MT is low, a higher prefetch distance is expected, and vice versa. (vii) Combine both prefetching and pinning. } 

\textcolor{black}{~\textbf{Generalizability:} Following the above Static Profiling Framework, we believe that memory-bound workloads (other than DLRMs) executing on GPUs can benefit from our key contributions with potential applications being Graph Neural Networks~\cite{song2023ugache} and Graph Mining~\cite{yuan2023everest}.}

~\textbf{Scalability and Industrial Adoption:} %Our findings suggest a significant performance improvement over off-the-shelf PyTorch.
Although we consider model sizes which can fit within one GPU, as our proposed techniques optimize the embedding table granularity, our solutions are applicable for large-scale distributed inference scenarios~\cite{matam2024quickupdate}. Further, the forward pass in the training pipeline~\cite{stash, arxiv_stash, skipper} could benefit from our schemes. By offering a readily deployable and performant solutions with prefetching and pinning, our work opens doors for wider industrial adoption of optimized DLRM inference pipelines.
%Thus, given our proposed scheme's plug-and-play nature facilitates seamless integration into existing workflows, we encourage its adoption by industry for high-performance deployments.
%Thus, given our plug-and-play designs, we encourage industrial adoption of our scheme for performant deployments.

%by following the three steps: (i) doing an in-depth characterization over the key metrics  (Table~\ref{tab:base_profiling}), (ii) evaluating the scope of optimal WLP (Figure~\ref{fig:vary_wlp}), and (iii) careful design space exploration with prefetching and cache pinning (Section~\ref{sec:optimizations}).

% Scalability of our work

% Adoption of our work with training

% By improving the irregular loads, we overall reduce batch's tail latency.

% Encourage adoption of our schemes in ML frameworks -- plug and play with quick software development. 

% what other applications can benefit from our proposed schemes? 

% How can our proposed schemes continue to work on top of future multi-threading efforts? 

%%%%%

\section{Conclusion}
With the ever-increasing compute and memory bandwidth requirements of DLRMs, they are increasingly getting adopted on GPUs. However, improving DLRM inference performance on GPUs needs co-examination of DLRM models and the underlying architectural artifacts. In this work, we show that the embedding stage continues to dominate the DLRM inference pipeline, causing a performance gap of up to 3.2$\times$ in the worst case. We show that standard embedding kernels underutilize the warp level parallelism (WLP) offered by the GPU hardware, and can be improved via compiler optimizations. Yet, the optimal  WLP is insufficient in fully hiding the long latency load stalls. To tackle this, we propose specialized techniques (software prefetching and L2 pinning), and also combine them.~\textcolor{black}{Without requiring any modifications in the hardware or models,} our experimental evaluations on A100 ~\textcolor{black}{and H100} GPU over large models and ~\textcolor{black}{a variety of datasets} indicate performance improvements by up to 103\% for the embedding stage, and up to 77\% for the overall inference. 
%We believe this establishes a new bar for any future research exploring hardware and software optimizations. 
~\textcolor{black}{We set a new benchmark for any future research, and believe that} our proposed designs can be generally applied to a wide range of memory-bound kernels.

%%%%% 

\section*{Acknowledgements}
\textcolor{black}{We extend our gratitude to the anonymous reviewers for their thorough feedback, which has significantly enhanced the paper through their valuable insights. This research was partially funded by NSF grants \{\#1931531, \#1955815, \#1763681, \#2116962, \#2122155, and \#2028929\}. We thank the NSF Chameleon Cloud project CHI-231143 for their generous compute grant, and extend special thanks to the members of HPCL. All product names used here are for identification purposes only and may be trademarks of their respective companies. 
%Finally, we wish to thank Aniruddha Vaidya for the stimulating intellectual discussion, Yaosheng Fu for guidance on L2 cache residency control, and Dr. Rui Zhang and Dr. Vijaykrishnan Narayan for providing access to A100 compute resources. 
}

% \section*{Acknowledgment}

% The preferred spelling of the word ``acknowledgment'' in America is without 
% an ``e'' after the ``g''. Avoid the stilted expression ``one of us (R. B. 
% G.) thanks $\ldots$''. Instead, try ``R. B. G. thanks$\ldots$''. Put sponsor 
% acknowledgments in the unnumbered footnote on the first page.

% \section*{References}

% Please number citations consecutively within brackets \cite{b1}. The 
% sentence punctuation follows the bracket \cite{b2}. Refer simply to the reference 
% number, as in \cite{b3}---do not use ``Ref. \cite{b3}'' or ``reference \cite{b3}'' except at 
% the beginning of a sentence: ``Reference \cite{b3} was the first $\ldots$''

% Number footnotes separately in superscripts. Place the actual footnote at 
% the bottom of the column in which it was cited. Do not put footnotes in the 
% abstract or reference list. Use letters for table footnotes.

% Unless there are six authors or more give all authors' names; do not use 
% ``et al.''. Papers that have not been published, even if they have been 
% submitted for publication, should be cited as ``unpublished'' \cite{b4}. Papers 
% that have been accepted for publication should be cited as ``in press'' \cite{b5}. 
% Capitalize only the first word in a paper title, except for proper nouns and 
% element symbols.

% For papers published in translation journals, please give the English 
% citation first, followed by the original foreign-language citation \cite{b6}.

% \begin{thebibliography}{00}
% \bibliography{refs}
% \end{thebibliography}
 \bibliographystyle{IEEEtran}
 %\bibliographystyle{plain}
 %\balance
 %\bibliography{refs}
 \bibliography{sample}  

 % LaTeX template for Artifact Evaluation V20240722
%
% Prepared by Grigori Fursin with contributions from Bruce Childers,
%   Michael Heroux, Michela Taufer and other colleagues.
%
% See examples of this Artifact Appendix in
%  * ASPLOS'24 "PyTorch 2: Faster Machine Learning Through Dynamic Python Bytecode Transformation and Graph Compilation": 
%      https://dl.acm.org/doi/10.1145/3620665.3640366
%  * SC'17 paper: https://dl.acm.org/citation.cfm?id=3126948
%  * CGO'17 paper: https://www.cl.cam.ac.uk/~sa614/papers/Software-Prefetching-CGO2017.pdf
%  * ACM ReQuEST-ASPLOS'18 paper: https://dl.acm.org/citation.cfm?doid=3229762.3229763
%
% (C)opyright 2014-2024 cTuning.org
%
% CC BY 4.0 license
%

% \documentclass{sigplanconf}

% \usepackage{hyperref}

% \begin{document}

%%%%%%%%%%%%%%%%%%%%%%%%%%%%%%%%%%%%%%%%%%%%%%%%%%%%
% When adding this appendix to your paper, 
% please remove above part
%%%%%%%%%%%%%%%%%%%%%%%%%%%%%%%%%%%%%%%%%%%%%%%%%%%%
\clearpage
\appendix
\section{Artifact Appendix}

%%%%%%%%%%%%%%%%%%%%%%%%%%%%%%%%%%%%%%%%%%%%%%%%%%%%%%%%%%%%%%%%%%%%%
\subsection{Abstract}

The artifact covers the complete steps to setup DLRM inference on GPUs. It provides the codebase for the proposed schemes: (1) improve WLP by lowering register allocation (2) various prefetching designs (3) L2 pinning design (4) most performant combined design (RPF + L2P + OptMT). Also, the necessary datasets are shared. Overall, the steps are shared to help reproduce the figures in the results section.

\subsection{Artifact check-list (meta-information)}

{\small
\begin{itemize}
  \item {\bf Algorithm: } DLRM inference
  \item {\bf Program: } DLRM implementation from Meta using PyTorch
  \item {\bf Compilation: } gcc 11.4.0, nvcc 12.2
  \item {\bf Model: } DLRM variants mentioned in Gupta et al – Section~\ref{sec:methodology}
  \item {\bf Data set: } Section~\ref{sec:methodology}
  \item {\bf Run-time environment: } Ubuntu 22.04.4 LTS
  \item {\bf Hardware: } CPU: AMD EPYC 7763 64-Core Processor, GPU: Nvidia A100-SXM4-80G (complete details in Table~\ref{tab:hardware}
  \item {\bf Metrics: } Batch Latency (ms), Speedup over base PyTorch
  \item {\bf Output: } Batch Latency (ms)
  \item {\bf Experiments: } Figure~\ref{fig:emb_only}
  \item {\bf How much disk space required (approximately)?: } 80GB
  \item {\bf How much time is needed to prepare workflow (approximately)?: } 2-3 hours
  \item {\bf How much time is needed to complete experiments (approximately)?: } Under 1 week
  \item {\bf Publicly available?: } Yes
\end{itemize}
}

%%%%%%%%%%%%%%%%%%%%%%%%%%%%%%%%%%%%%%%%%%%%%%%%%%%%%%%%%%%%%%%%%%%%%
\subsection{Description}

\subsubsection{How to access}
The codebase is available on Zenodo at \url{https://doi.org/10.5281/zenodo.13325108} and Github at \url{https://github.com/rishucoding/reproduce_MICRO24_GPU_DLRM_inference}

\subsubsection{Software dependencies} The required software dependencies are outlined in the repository. 

\subsubsection{Data sets} The required dataset files are added to the repository.

\subsubsection{Models} Steps to save and load models are added in the repository.

%%%%%%%%%%%%%%%%%%%%%%%%%%%%%%%%%%%%%%%%%%%%%%%%%%%%%%%%%%%%%%%%%%%%%
\subsection{Installation}
The detailed installation steps are mentioned in the artifact, and the following is a high-level summary of the steps: 
\begin{enumerate}
    \item Install Anaconda
    \item Install PyTorch
    \item Evaluate baseline performance over various datasets.
    \item Evaluate OptMT over various datasets.
    \item Evaluate RPF+OptMT over various datasets.
    \item Evaluate L2P+OptMT over various datasets.
    \item Evaluate RPF+L2P+OptMT over various datasets.
\end{enumerate}

%%%%%%%%%%%%%%%%%%%%%%%%%%%%%%%%%%%%%%%%%%%%%%%%%%%%%%%%%%%%%%%%%%%%%
\subsection{Experiment workflow}
We suggest to follow the README.md file in the above repository. 
%%%%%%%%%%%%%%%%%%%%%%%%%%%%%%%%%%%%%%%%%%%%%%%%%%%%%%%%%%%%%%%%%%%%%
\subsection{Evaluation and expected results}
Figure~\ref{fig:emb_only} can be directly reproduced following the given steps.

%%%%%%%%%%%%%%%%%%%%%%%%%%%%%%%%%%%%%%%%%%%%%%%%%%%%%%%%%%%%%%%%%%%%%
\subsection{Experiment customization}
Models, datasets, and optimization designs can be customized to evaluate various configurations, and thus reproduce majority of results shown in Section~\ref{sec:evaluation}.
%%%%%%%%%%%%%%%%%%%%%%%%%%%%%%%%%%%%%%%%%%%%%%%%%%%%%%%%%%%%%%%%%%%%%
\subsection{Notes}
Please raise a Github issue for any questions. 
%%%%%%%%%%%%%%%%%%%%%%%%%%%%%%%%%%%%%%%%%%%%%%%%%%%%%%%%%%%%%%%%%%%%%

%%%%%%%%%%%%%%%%%%%%%%%%%%%%%%%%%%%%%%%%%%%%%%%%%%%%
% When adding this appendix to your paper, 
% please remove below part
%%%%%%%%%%%%%%%%%%%%%%%%%%%%%%%%%%%%%%%%%%%%%%%%%%%%

% \end{document}

\end{document}